\documentclass[twocolumn]{aastex6}

\usepackage{color}
\usepackage{amsmath}  

\begin{document}

\title{Optical properties of non-stoichiometric amorphous silicates with application to circumstellar dust extinction }

\author{Akemi Tamanai\altaffilmark{1} and Annemarie Pucci\altaffilmark{1}}
\affil{Universit\"at Heidelberg, Kirchhoff-Institut f\"ur Physik,
           Im Neuenheimer Feld 227,
           69120 Heidelberg, Germany
}

\author{Ralf Dohmen\altaffilmark{2}}
\affil{Ruhr Universit\"at Bochum, Institut f\"ur Geologie, Mineralogie und Geophysik,\\
           Universit\"atsstr. 150,
           44780 Bochum, Germany           
}

\author{Hans-Peter Gail\altaffilmark{3}}
\affil{Universit\"at Heidelberg, Zentrum f\"ur Astronomie, 
           Institut f\"ur Theoretische Astrophysik,\\
           Albert-Ueberle-Str. 2,
           69120 Heidelberg, Germany 
}

\begin{abstract}
We determine the optical constants of non-stoichiometric amorphous magnesium-iron silicates and demonstrate that these can well reproduce the observed mid-infrared emission spectra of evolved stars. Stoichiometric and non-stoichiometric amorphous magnesium-iron silicate films are fabricated by pulsed laser deposition. Transmittance and ellipsometry measurements are performed in the wavelength range between 2 and 200 $\mu$m and 1.7 and 33 $\mu$m, respectively. Optical constants are derived from transmittance and ellipsometric $\Psi$ and $\Delta$ spectra by means of oscillator models. These newly obtained optical constants are applied in radiative transfer models for examining reproducibility of the observed spectral features of circumstellar dust shells around supergiants. The spectra of four selected supergiants are dominated by amorphous silicate dust emission in the wavelength range of 9 and $25\ \mu\mathrm{m}$. To obtain a good fit to the observed spectra, we take into account amorphous corundum and metallic iron particles as additional dust components to the model calculations to rationalize the dust emission at $\lambda<8\ \mu$m. For each of the objects, a set of model parameters (dust mass, condensation temperature) is derived by an automated optimization procedure which well reproduces the observation. Consequently, our model spectra using new optical data find that the silicate bands at $\sim10$ and $\sim18\ \mu$m depend on the magnesium and iron ratio in the silicate system, and that a good fit requires a significant iron content of the amorphous silicate dust component to reproduce the observed peak positions and shape of the silicate bands. 
\end{abstract}

\keywords{circumstellar matter -- stars:  mass-loss -- stars:  winds, outflows
--  stars: AGB and post-AGB }

\maketitle

\section{Introduction}

The standard cosmic element mixture found almost everywhere in space is oxygen-rich. The oxygen abundance exceeds that of carbon usually by a factor of about two \citep[see, e.g.,][]{Asp09,Lod09}, in which case carbon is chemically blocked in the CO molecule. If refractory minerals form from such matter the resulting mixture of condensed phases is strongly dominated by the magnesium-iron (hereafter Mg-Fe) silicates \citep[e.g.][]{Gai10}. Silicate dust is therefore ubiquitous in space if conditions are such that solids can resist to the ever-present destroying mechanisms. Mg-Fe silicates form the dominating dust component in circumstellar environments like protostellar accretion disks or dust shells of evolved late type stars (if these are not carbon-rich), and forms, besides carbon dust, the main dust component of interstellar matter. 

The nature of the dust material in space is inferred from the wavelength dependence of the interstellar extinction or the infrared (IR) emission spectrum from circumstellar environments. In particular some strong IR bands from lattice vibrations are indicative of the nature of the material present and these have been extensively used for remote sensing of the dust composition in space \citep[see][ for a review on observations]{Mol10}. For silicate dust one uses in particular their absorption bands from stretching and bending vibrations of the SiO$_4$-tetrahedron around 10~$\mu$m and 18~$\mu$m, respectively, and if the silicates are
crystalline, there are many additional absorption bands up to 100~$\mu$m \citep[e.g.][]{Koi03} that can be observed.

Since the earliest days of research on dust in space, it is known \citep[see][ for a historical review]{Li05} that the interstellar silicate dust component has an amorphous lattice structure. This is recognized by the broad and smooth shape of the IR bands around 9.7~$\mu$m and 18~$\mu$m without showing the characteristic sub-structures seen for crystalline silicates, and by the absence of far-infrared (FIR) bands $\gtrsim$~25~$\mu$m of crystalline materials. Crystalline silicate dust also has been found in space, but this is limited to special environments: circumstellar dust shells and protostellar accretion disks \citep{Mol10}. While for accretion disks, it is quite common to have part of the silicate dust in crystalline form and most of it as amorphous materials \citep[e.g.][]{Oli11}, for circumstellar dust shells the presence of some fraction of crystalline silicate dust is restricted to an only small fraction of the objects \citep[cf.][]{Jon12}. Most part of the silicate dust in accretion disks and circumstellar envelopes is seen to have an amorphous lattice structure like interstellar medium (ISM) dust.

The analysis of the observed dust features requires a detailed knowledge of the optical properties which are largely ruled by the complex dielectric function $\varepsilon$($\omega$) of the dust grains. Since silicate dust in space mostly has an amorphous structure, much effort has been undertaken to determine the variation of $\varepsilon$($\omega$) for amorphous silicates with different compositions and prepared by various fabrication techniques. \citet{Col03} have reviewed the experimental work and its main results some time ago. 

Optical data of olivine-like amorphous silicates with different Fe-content are of special importance for astronomical applications. While crystalline silicate dust formed from a cosmic element mixture under near equilibrium conditions, which are a mixture of almost Fe-free forsterite and enstatite \citep[e.g.][]{Gro72,Gai10}, the amorphous cosmic dust seems to have a different composition due to strongly deviating formation conditions in non-equilibrium condensation processes. Interstellar silicate dust seems to be dominated by an amorphous material with composition approximately equal to MgFeSiO$_4$  \citep{Dra03}. In accretion disks, the amorphous dust component seems to be a more complex mixture of materials with olivine-like (Mg$_{2x}$Fe$_{2(1-x)}$SiO$_4$) and pyroxene-like (Mg$_{x}$Fe$_{1-x}$SiO$_3$) composition with a considerable iron content ($1-x\approx0.3$), \citep[see][]{Pol94}, but also high fractions of crystalline olivine and pyroxene have been reported in some cases \citep[e.g.][]{Wat09}. In circumstellar dust shells, the amorphous dust component seems also to consist of olivine- and pyroxene-like materials with varying iron content \citep{Mol10}.

Laboratory-measured optical properties of various amorphous silicates have been reported so far. \citet{Day79,Day81} fabricated amorphous Mg-~(MgSiO$_3$ $\&$ Mg$_2$SiO$_4$) and Fe-~(FeSiO$_3$ $\&$ Fe$_2$SiO$_4$) silicate thin films on KBr (potassium bromide) disks by means of reactive sputtering for transmission measurements in IR regions. Optical constants of these films were derived by use of classical dispersion theory. Amorphous silicate glasses with different proportions of Fe and Mg in both pyroxene- and olivine-like systems were produced by making use of quenching-melts technique, and both transmittance and reflectance measurements were carried out with KBr pellet and thin slab methods, respectively, in the wavelength range between 0.19 and 500~$\mu$m \citep{Dor95}. They derived the optical constants of each amorphous silicate glass by Kramers-Kronig relations. Likewise, \citet{Jae03} adopted the sol-gel technique so as to fabricate non-stoichiometric pyroxene samples of the composition Mg$_x$SiO$_y$ (x=0.7, 1, 1.5, 2, 2.4; y=2.7, 3, 3.5, 4, 4.4), 
and reflectance spectra of dense pellets embedded in epoxy resin and transmittance spectra of KBr (for MIR) and PE (polyethylene for FIR) were measured as functions of wavelength in the range 0.2 -- 200~$\mu$m. These optical constants were obtained by either the Lorenz-oscillator fit method or Kramers-Kronig relations or both. In most cases, experimentally measured optical properties are represented by absorption, transmission, or reflection \citep{Koi87,Tam06,Spe11}. In fact, highly accurate optical constants of various materials are an essential prerequisite for astrophysical modelings e.g.~interstellar dust, accretion disks around young stars, planetary atmospheres, and cometary dust tails. However, optical constants of diverse chemical compositions, especially, amorphous materials with different conditions such as temperature- and pressure-dependent (e.g.~high temperature $\gtrsim1273~K$) and morphological aspect including particle irregularity in shape, aggregate/agglomerate states, and various individual particle sizes are not that much available.

In this paper, we describe the optical properties of non-stoichiometric amorphous silicates with composition between olivine-like and pyroxene-like and varying iron content, fabricated by pulsed laser deposition (PLD) technique in the MIR and FIR regions so as to characterize the effect of metals (Mg $\&$ Fe) on the silicate absorption. For cross-checking the optical constants of each investigated silicate, we carried out both spectroscopic transmittance and ellipsometry measurements. As an application, the newly derived optical constants are utilized to compare synthetic spectra of dust shells around supergiants to observed MIR spectra of four objects.

\section{Silicate stardust}
\label{SectPresolar}

Information on the composition and size of silicate dust grains from circumstellar dust shells enshrouding highly evolved stars can be obtained in two ways. 

The standard way is to compare IR spectra from the objects of interest with template spectra calculated from measured optical properties of laboratory made analogous minerals and with models for the source regions (see, e.g., \citet{Mol10}, \citet{Hen10}, and \citet{Col03} for reviews).

The second possibility is to analyze specimens of cosmic dust in the laboratory. This has become possible by the detection in the 1980's that primitive meteorites contain a small fraction of stardust grains of different composition from different stellar sources (see, e.g., the collection of articles in \citet{Ber97} or the review of \citet{Lod05}), the presolar dust grains, which are identifyable by their suspicious isotopic anomalies found for some elements. This allows a direct study of their properties and composition in the laboratory. Since the detection by \citet{Ngu04} that also silicates from stellar sources are present in meteorites it has become possible also to analyze presolar silicate dust grains. \citet{Vol09a}, \citet{Bos10}, \citet{Ngu10}, and \citet{Bos12} have provided direct information on the size and composition for 145 such presolar silicate grains. 

We concentrate here on silicate grains from the meteorite ACFER 094, a primitive ungrouped carbonaceous chondrite, because the matrix material in this meteorite seems to be only slightly metamorphosed. The matrix material has not been subject to high temperatures where the material would equilibrate and the presolar silicate grains in this meteorite show only indications of minor aqueous alteration and oxidation of Fe metal on the parent body \citep[e.g.][]{Kel12}. Hence, the compositions of the silicate grains probably largely resemble their composition at formation time in the dust shell of an AGB star.

The fractional abundance of the six major rock forming elements (O, Si, Mg, Fe, Ca, and Al) was studied in  \citet{Vol09a}, \citet{Bos10}, \citet{Ngu10}, and \citet{Bos12}. We try to derive from the laboratory derived fractional abundances of the elements given in these papers the corresponding abundances of the oxides SiO$_2$, SiO, MgO, FeO, CaO and Al$_2$O$_3$ and of solid iron in the dust grains which reproduces their observed elemental composition and considers also the stiochiometry of the elements.

\begin{figure}[t]

\includegraphics[width=.47\hsize]{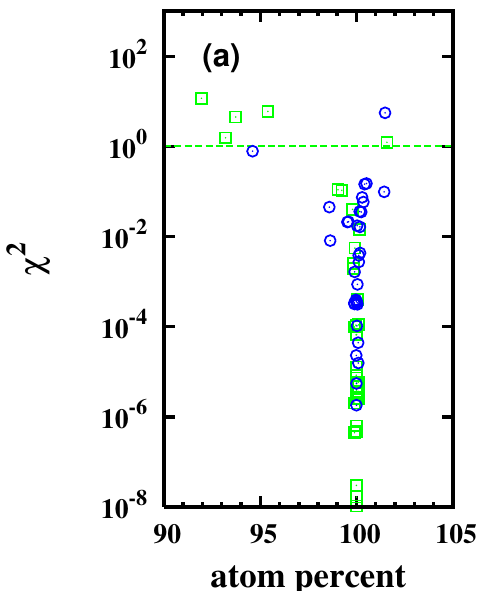}
\hfill
\includegraphics[width=.49\hsize]{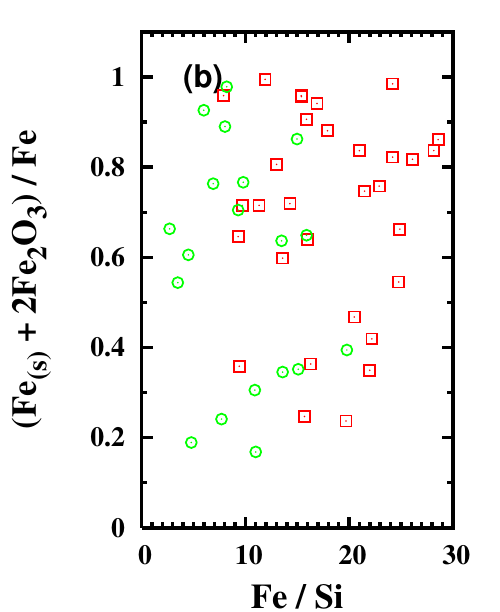}

\caption{(a) Least square fit of composition of presolar silicate grains from meteorite ACFER 094. The value of $\chi^2$ is plotted versus the sum of all atom-percents of the optimum fit of the elements. Circles correspond to data  from \citet{Vol09a}, squares to data from \citet{Bos10}. (b) Fraction of the total Fe not bound in the silicates, plotted for all grains for which $\chi^2<0.01$, versus the atom percent of Fe in the grains. $Fe_{(s)}$ denotes the solid iron component.
\label{FigStoichFit}
}

\end{figure}

The oxide SiO is added to the set because there exist complex silicate structures (e.g. chains of rings) where O/Si ratios formally correspond to an oxide component SiO$_x$ with $1\le x\le2$ such that the astronomical silicates may deviate in their composition from the usually assumed olivine or enstatite stoichiometry. To account for a more complex stoichiometry the SiO is included.

We add metallic iron to the set of oxide components since it is known that presolar silicate dust grains may contain iron grains as nm-sized inclusions, at least in some cases. Also  iron sulphide particles are commonly found as nm-sized inclusions if iron particles are present. These inclusions are found in GEMS, a class of subgrains in interplanetary dust particles (IDPs) which are characterized by the presence of iron and iron-sulphide inclusion embedded in a glassy ground matrix. While most of such grains seem to be of interstellar or solar system origin, a few percent show oxygen isotopic anomalies pointing to an origin as stellar or supernova condensate \citep[e.g.][]{Mes03,Mat08,Kel11}.
We cannot take account of the presence of FeS particles because no S abundance was measured. 

Additionally we consider Fe$_2$O$_3$ as a possible component. It is unlikely that iron can be oxidized to the Fe$^{3+}$ state under the strongly reducing circumstellar conditions, but the presolar material may be oxidized during its residence on the parent body of the meteorite or on the earth's surface before being collected. 

\begin{figure}[t]

\includegraphics[width=\hsize]{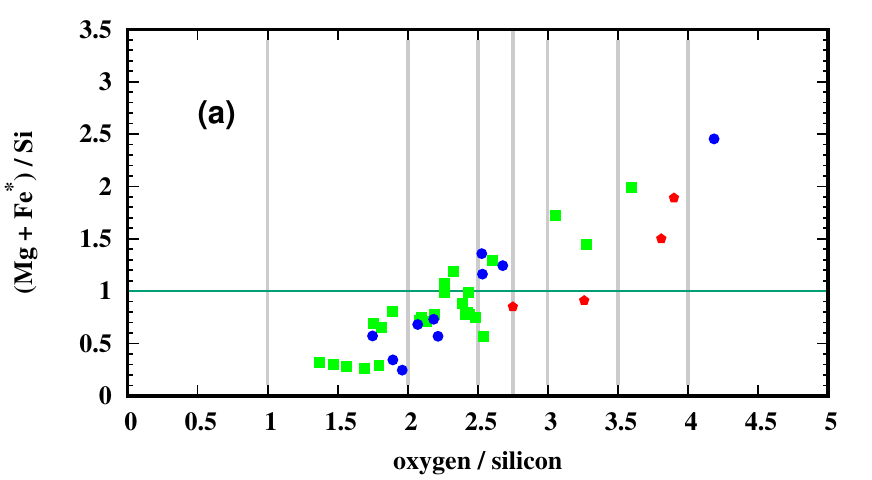}
\includegraphics[width=\hsize]{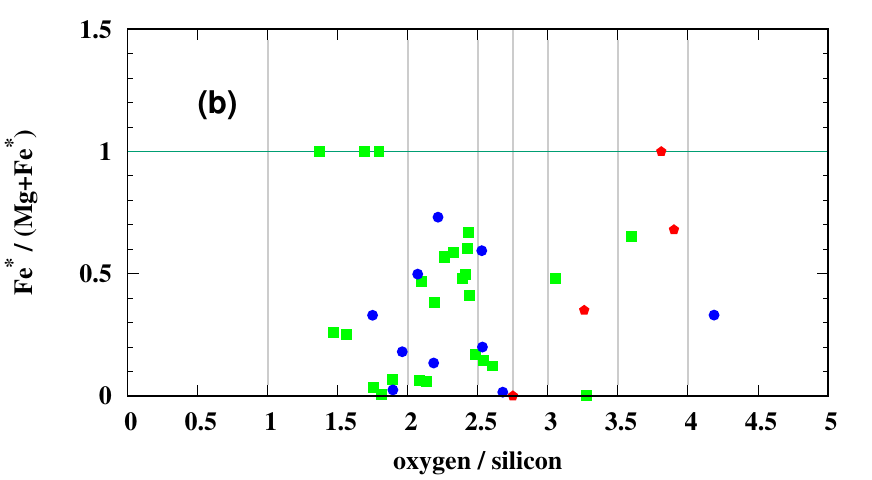}

\caption{Composition of presolar silicate grains from meteorite ACFER 094. Circles correspond to data  from \citet{Vol09a}, squares to data from \citet{Bos10}. The red diamonds refer to the synthetic amorphous silicates studied in this paper (see Table \ref{table: 1}). The Fe$^*$ refers to that part of the total iron content that is part of the silicate lattice. {\bf(a)} Correlation between cation/Si ratio and O/Si ratio. The gray vertical bars indicate the O/Si ratio for a number silicate mineral structures and of solid SiO. {\bf(b)} Number of  Fe$^*$ atoms in the silicate lattice relative to total number of cations (Mg+Fe$^*$).}
\label{FigStoichSil}
\end{figure}

This set of components does not cover all possibilities for the composition of presolar material from oxygen-rich parent stars, but the most important ones. The relative abundances of the components are determined by a least square minimization such that the relative abundance of the sum of all elements in the mixture fits as well as possible the measured elemental abundances. Figure \ref{FigStoichFit}a shows the resulting value of $\chi^2$ as function of the fractional abundance of the sum of all elements contained in the particle (in the figure: the percentage). Ideally, this should sum up to 1.0 or to 100\%. For most particles a solution can be found breaking down the material into the fundamental oxides and the additional Fe that fits the experimental data with an accuracy better than $10^{-2}$; in many cases even much better. This suggests the corresponding particle likely to be composed of minerals (and free iron) that contain the seven oxide components SiO$_2$, SiO, MgO, FeO, Fe$_2$O$_3$, CaO, Al$_2$O$_3$. For particles where the optimum fit fails to reproduce the measured element abundances within the limits of $10^{-2}$ the grains either contain significant fractions of additional elements, or components with an alternate composition, or the measurements are inaccurate for some reasons. We exclude those particles from further considerations.

\begin{table*}
\caption{RBS analysis of amorphous magnesium-iron silicates. }
\label{table: 1}
\centering
\normalsize
\begin{tabular}{c c c c c c}	
\hline
\hline
\noalign{\smallskip}
Sample & Film        & Target      & Thin film     & Mg/(Fe+Mg)    & (Mg+Fe)/Si   \\
       &  Thickness  & composition & properties    &  &  \\
       &   [nm]      &             & Chem. Formula &  &  \\
\noalign{\smallskip}
\hline
\noalign{\smallskip}
Si-film1    & 235    & Fo100    & Mg$_{0.85}$SiO$_{2.75}$            & 1.00 & 0.85 \\
\noalign{\smallskip}
Si-film2    & 210    & Fo80Fa20 & Mg$_{0.58}$Fe$_{0.33}$SiO$_{3.26}$ & 0.64 & 0.91 \\
\noalign{\smallskip}
Si-film3    & 235    & Fo40Fa60 & Mg$_{0.60}$Fe$_{1.29}$SiO$_{3.9}$  & 0.32 & 1.89 \\
\noalign{\smallskip}
Si-film4    & 70     & Fa100    & Fe$_{1.50}$SiO$_{3.81}$            & 0.00 & 1.50 \\
\noalign{\smallskip}
\hline
\end{tabular}

\end{table*}

With respect to iron-bearing components found in the fit the following holds: Metallic iron must be an inclusion in the silicate or attached to its surface, and any portion of Fe$_2$O$_3$ must be a secondary product of oxidation of iron metal during the residence time on the parent body or on earth. Only the fraction of Fe not contained in these two components could be part of the silicate material. Figure \ref{FigStoichFit}b shows the fraction of the total Fe content that seems to have been built into the presolar silicate grains as iron metal during particle formation. Surprisingly, the laboratory determined composition of the silicate grains can be best explained if a high fraction of the Fe contained in the presolar grains is not part of the silicate lattice but is contained as metallic iron particles, either as inclusions within the grains or attached to its surface. The size of such iron particles must be generally below the resolution limit of the nano-SIMS ($\sim$~50~nm) used for the investigations  \citep{Vol09a,Bos10, Ngu10,Bos12} because it is not reported in the papers that such inclusions are detected in all or most of the grains. Atom ratios Fe/Si up to 0.3 are frequent, which convert to metal/silicate volume ratios of up to about 0.15.

Part of the particles contain some fraction of Al and Ca. This indicates heterogeneous composed particles of silicate and aluminum-calcium minerals or a complex chemical composition of the grains. To avoid any ambiguity resulting from this we do not consider Ca-Al-bearing particles. The remaining particles are (more or less) pure magnesium-iron-silicate particles where part of the iron seems to be present as metallic iron.

Figure \ref{FigStoichSil}a shows the correlation between the cation to silicon ratio in the presolar grains. The vertical gray lines correspond to the O/Si ratios in different mineral structures varying from O/Si = 2 for framework silicates to O/Si = 4 for island silicates and intermediate values for chain and ring silicates, and for solid SiO (O/Si = 1). The presolar silicates cover the whole range with no  preference for one of the specific values corresponding to well-defined mineral structures. This suggests the amorphous character of circumstellar silicates does not solely refers to a disorder in the arrangement of the SiO$_4$-building blocks in a material with otherwise well defined composition as for mineral compounds, as it was obtained by melt-quenching \citep{Dor95}. Such kind of disorder will also manifests itself in broad smeared-out $\sim10\ \mu$m and $\sim18\ \mu$m silicate bands. The silicates condensed in stars seem to have more than that also compositions with highly variable fractions of bridging bonds between the SiO$_4$-tetrahedrons not corresponding to any of the regular structures of minerals. They may resemble the ``chaotic silicates'' discussed by \citet{Nut90} and such a kind of material may also form during the rapid quenching of vapor by the vapor deposition fabrication method used for preparing the amorphous thin films used by us (see Sect.~\ref{SectPrepSamp}). 

Figure \ref{FigStoichSil}b shows the iron content of the silicate material according to our calculation. Significant iron contents seem to be frequent, but also iron-poor silicates seem not to be rare. Probably this means that the iron fraction of silicates formed in the outflow from stars is variable, but since the presolar grains are a mix from many different stellar sources, one cannot draw any conclusion on the circumstances which determine the iron content. Additionally one has to be aware that our calculated FeO fraction may contain some contributions from rusting of iron, if also Fe$_2$O$_3$ is found to have non-negligible abundance.

\section{Experiments}

\subsection{Non-Stoichiometric Amorphous Silicate Samples}
\label{SectPrepSamp}

Non-stoichiometric amorphous samples were prepared by means of PLD technique at the Institute of Geology, Mineralogy, and Geophysics of the Ruhr-University Bochum (RUB) \citep[e.g.][]{Doh02}. Floating zone Si (111) crystal structure substrates with size of $10\times10$~mm and thickness of 740~$\mu$m (SILTRONIX) were employed since Si wafer is highly transparent via IR region. 

An Excimer-Laser with nanosecond pulses of 193~nm wavelength at a frequency 10~Hz was utilized for ablating a synthesized target pellet material under a high vacuum condition (10$^{-5}$~mbar) in the PLD chamber. The laser fluences were between 1 and 5~J cm$^{-2}$, which yielded typically a film thickness of about 50~nm after 10 minutes. The different stoichiometry of each olivine target pellet was prepared by mixing thoroughly three synthetic powders which were 99.99$\%$ pure SiO$_2$, MgO, and Fe$_2$O$_3$ with the nominal ratio in a mortar, and this mixed powder was cold-pressed into a pellet of 8 mm diameter. Thereupon, the fabricated target pellets were annealed in a furnace for approximately 20 hours at temperatures of 1500, 1300, 1225, 1225, and 1100~$^{\circ}$C under oxygen fugacity $f_{\rm O_2}$ of 1, 10$^{-9}$, 10$^{-10}$, 10$^{-11}$, and 10$^{-12}$~bar for Mg$_2$SiO$_4$, Mg$_{0.8}$Fe$_{1.2}$SiO$_4$, Mg$_{0.4}$Fe$_{1.6}$SiO$_4$, and Fe$_2$SiO$_4$, respectively. The $f$O$_2$ was controlled by the flow of gas mixture (CO and CO$_2$) during the annealing processes \citep[e.g.][]{Doh02}. The T--$f$O$_2$ conditions were selected for the Fe-bearing pellets such as to reduce the Fe$^{+3}$ to Fe$^{+2}$ and to be within the stability field of the respective olivine composition according to \citet{Nit74}.

In the PLD chamber, the $10\times10$~mm Si substrate was placed on a substrate holder that was mounted directly in front of the target material on a rotating holder. This Si substrate was first heated up to 400$^{\circ}$C under a pressure of 10$^{-5}$~mbar for 15 minutes in order to remove unnecessary volatile absorbents on the surface, especially H$_2$O. The deposition was carried out approximately at room temperature and a background gas pressure of around 10$^{-6}$~mbar. When an incoming laser beam hits the target, a plasma cloud containing a mixture of neutral and ionic atoms was created at an anterior direction of the mounted Si substrate. Hence, the particles in the plasma are deposited on the mounted Si substrate (more details in \citet{Doh02} and Tamanai et al., in prep.).

The chemical compositions of the deposited silicate films were determined by Rutherford Backscattering spectroscopy (RBS). The RBS measurements were carried out at the RUBION facility of the RUB using the Dynamitron Tandem accelerator to produce a 2~MeV beam of alpha particles. The measurements were performed with a final aperture of 1~mm, a beam current typically between 20 and 50~nA, a detecting angle of 160$^{\circ}$ with a silicon particle detector at an energy resolution of about 16 -- 20~keV, and the sample surface was tilted at about 5$^{\circ}$ relative to the beam direction to avoid channelling. The RBS spectra were simulated by using the software RBX \citep[Version 5.18 ][]{Kot94} assuming a density of the amorphous silicate layers as of the respective crystalline olivine. The thin film thickness was not completely homogeneous over the sample; it was a maximum for the central area. Uncertainties are $\pm5$~nm minimum related to the energy resolution of the detector. However, the fabricated silicate films are chemically homogeneous at least down to the 10~nm scale and total amorphous structure which has been demonstrated by TEM investigations of thin films with similar compositions \citep{Doh02,LeG15}. The oxidation state of Fe is mainly 2+ as shown by \citet{LeG15}. Table~\ref{table: 1} shows the physical and chemical characteristics of the deposited amorphous silicate samples analyzed by the RBS. 

\begin{figure}[t]

\includegraphics[width=1\hsize]{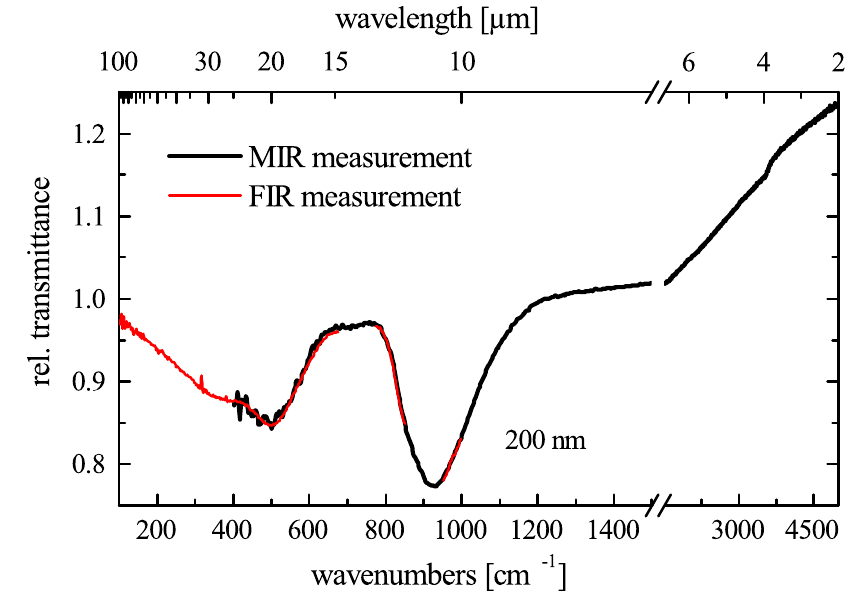}

\caption{Relative transmittance measurement of Si-film3 sample in the MIR and FIR region.}
\label{fig:Fo40T} 
\end{figure}

\subsection{FTIR Spectroscopy}
\label{SectFTIR}

The IR spectroscopic transmittance measurements have been performed in order to examine the optical properties of the deposited amorphous silicates. A Bruker IFS66~v/S Fourier transform infrared (FTIR) spectrometer with deuterium triglycine sulfate (DTGS) detector was employed for the MIR transmittance measurements in the frequency range between 500 and 5000~cm$^{-1}$ (20 -- 2~$\mu$m) as well as a Bruker Vertex80v with a special FIR-DTGS detector together with a Mylar beam splitter for the FIR measurements in frequency range between 50 -- 600~cm$^{-1}$ (200 -- 16.7~$\mu$m). The sample compartment of both FTIR spectrometers was evacuated under a pressure below 4~mbar for all measurements. All transmittance spectra were normalized to the spectrum of the bare Si(111) wafer and obtained with a spectral resolution of 4~cm$^{-1}$. As shown in Figure.~\ref{fig:Fo40T}, the transmittance spectra measured in both MIR and FIR regions overlapped each other at around 500~cm$^{-1}$ and were feasible to connect (more details in \citealt{Wet12} and \citealt{Wetzel12c}). Hence, it is possible to determine the optical constants through the incorporated transmittance spectra that can cover a wide spectral range within the IR region. 

\begin{figure}[t]

\includegraphics[width=1\hsize]{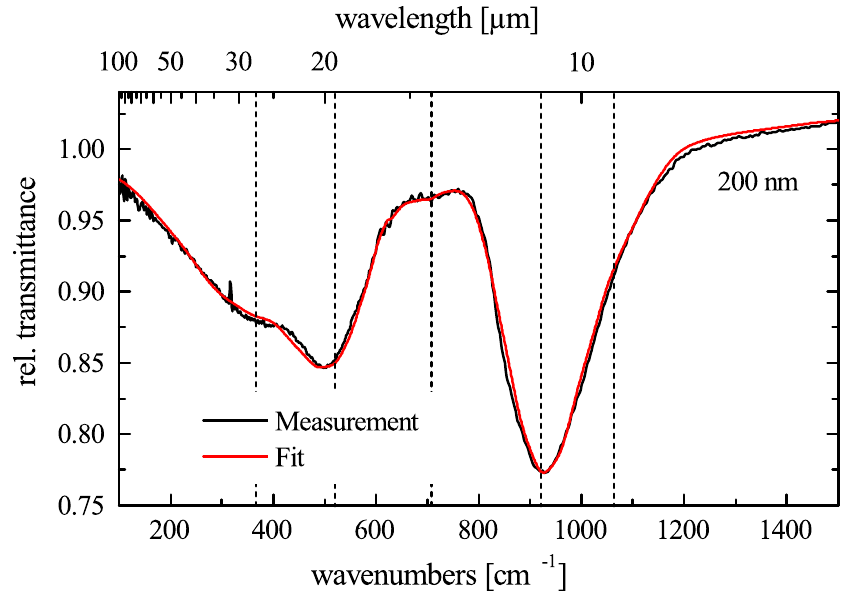}

\caption{Comparison of composite transmission spectrum of Si-film3 with the best fit of a Brendel oscillator model with the oscillator parameters given in Table~\ref{table: 2}. The oscillator positions are indicated by dashed lines.
}
\label{fig:Fo40C} 
\end{figure}

The measured transmittance spectrum of Si-film3 together with a Brendel oscillator model fit \citep{Brendel92} for the dielectric function
\begin{align}
\varepsilon(\omega)_{\rm IR}&=\varepsilon_\infty+\displaystyle\sum_{j=1}^{N}\frac{1}{\sqrt{2\pi}\sigma_{j}}\times
\nonumber\\
&\int_{-\infty}^{\infty}{\rm dz}~{\rm e}^\frac{-(z-\omega_{0,j})^2}{2\sigma^2_j}~\frac{\omega^2_{{\rm p},j}}{{\rm z}^2-\omega^2+{\rm i}\gamma_j\omega}
\label{EqBrendOscMod}
\end{align}
using five oscillators is shown in Figure~\ref{fig:Fo40C}. This model accounts for the amorphous structure of a material by assuming that the different IR modes can be represented by Lorentz-oscillators with randomly shifted resonance frequencies distributed according to Gaussian probability distributions. We performed the spectral fits in the range between 100 and 1400~cm$^{-1}$ using the software package SCOUT \citep{Scout2011}. For modeling the transmittance spectrum, the value of $\varepsilon_{\infty}$ has to be used as an input parameter. However, no values for $\varepsilon_{\infty}$  obtained from amorphous silicates equivalent to our non-stoichiometric samples are available.

Fundamentally, the optical properties are sensitively influenced by chemical compositions \citep{Koi03,Tam09}, and thus the value of $\varepsilon_{\infty}$ also varies with different concentrations of Fe and Mg in the silicate system. Consequently, the values of high frequency dielectric constant ($\varepsilon_{\infty}$) derived from the ellipsometric fitting have been applied for the transmittance modeling in order to determine the accurate dielectric function (see Sect.~\ref{SectEllips}). Since the oscillator damping constant $\gamma$ cannot properly be determined from the fits for vibrational spectra of disordered solids \citep[cf.][]{Ishikawa00}, we fixed the value of $\gamma$ to the resolution of our measurement 4~cm$^{-1}$. Nearly the same kind of procedure is already reported in the literature \citep{Ishikawa00,Brendel92, Grosse86,Naiman85}. Our parameters of the best fit for the oscillator model are given in Table~\ref{table: 2}. The Brendel oscillator model provides us a reasonable fit to the experimental transmittance spectra of our amorphous silicate samples.

\begin{table}
\caption{Parameters for the Brendel oscillators. }
\label{table: 2}
\centering
\normalsize
\begin{tabular}{c c c c c c }	
\hline
\hline
\noalign{\smallskip}
Osc.      & Param.                          & Si-film1                  &  Si-film2            &  Si-film3              & Si-film4           \\
             &  [cm$^{-1}$]                        &                         &                     &                      &                   \\
\noalign{\smallskip}
\hline
\noalign{\smallskip}
1           & $\omega_0$                   &  377                 &   372            &   366             &  307            \\
             & $\omega_p$                   &  514	               &    625            &   614             &  391            \\
						 & $\sigma$                              &  125                &    139            &   135             &   94              \\[.2cm]
2           & $\omega_0$                   &  502                 &    525           &    520            &   493             \\
             & $\omega_p$                   &  390                 &    410           &    401            &   452             \\
						 & $\sigma$                              &  75                   &    73             &    61              &    61              \\[.2cm]
3           & $\omega_0$                   &  739                 &   721            &     708           &    687            \\
             & $\omega_p$                   &  223                 &   181            &     191           &   174             \\
						 & $\sigma$                              &  70                   &    42             &    56              &    74              \\[.2cm]
4           & $\omega_0$                   &  972                 &   942            &     921           &    920            \\
             & $\omega_p$                   &  692                 &   752            &     670           &    628            \\
					   & $\sigma$                              &  77                   &    77             &     67             &    62              \\[.2cm]
5           & $\omega_0$                   & 1118                &  1098           &    1064          &    1062          \\
             & $\omega_p$                   &  212                 &   299            &     268           &    396            \\
					   & $\sigma$                              &  63                   &    74             &     58             &    79              \\
\noalign{\smallskip}
\hline
\noalign{\smallskip}
$\varepsilon_{\infty}$ &                    & 2.43                 &    2.67          &    3.15           &    3.69           \\
\noalign{\smallskip}
\hline
\end{tabular}
\tablecomments{
The values of $\varepsilon_{\infty}$ are obtained from the IRSE modeling.}

\end{table}

\subsection{Infrared Spectroscopic Ellipsometry}
\label{SectEllips}

\begin{table}

\caption{Parameters for ellipsometric modeling (Gaussian). }
\label{table: 3}
\centering
\normalsize
\begin{tabular}{c c c c c c }	
\hline
\hline
\noalign{\smallskip}
Osc.        & Param.             & Si-film1   &  Si-film2  &  Si-film3  & Si-film4  \\
            &  [cm$^{-1}$]       &            &            &            &                   \\
\noalign{\smallskip}
\hline
\noalign{\smallskip}
1		& {$\omega_0$}     &            &            &   369      &            \\
            & {$A$}            &            &            &   4.36     &            \\
            & {$\sigma$}       &            &            &   139      &            
\\[.2cm]
2		& {$\omega_0$}     &  461       &    448     &    501     &   412      \\
            & {$A$}            &  3.83      &   4.51     &   1.66     &   4.61     \\
            & {$\sigma$}       &  96	  &    106     &    65      &   134      
\\[.2cm]
3		& {$\omega_0$}     &  776       &    751     &    746     &    910     \\
            & {$A$}            &  0.72      &    0.44    &    0.59    &   2.57     \\
            & {$\sigma$}       &  39        &    26      &    20      &   54      
\\[.2cm]
4		& {$\omega_0$}     &  964       &   931      &    903     &    981     \\ 
            & {$A$}            &  3.62      &   3.61     &     3.43   &    2.09    \\
            & {$\sigma$}       &  87        &   88       &     70     &    123     
\\[.2cm]
5		& {$\omega_0$}     & 1157       &    1129    &     1021   &            \\
            & {$A$}            &  0.36      &   0.29     &     0.76   &            \\
            & {$\sigma$}       &  47        &   56       &     77     &            
\\
\noalign{\smallskip}
\hline
\noalign{\smallskip}
$\varepsilon_{\infty}$ &       & 2.43       &    2.67    &    3.15    &    3.69           \\
\noalign{\smallskip}
\hline
\end{tabular}

\end{table}
%

For obtaining the real and the imaginary part of the optical constants of the amorphous silicate samples, we carried out infrared spectroscopic ellipsometry (IRSE) as well. IRSE is a characterization technique for obtaining physical (e.g.~thickness and surface roughness) and optical properties of both anisotropic and isotropic materials. Fundamentally, ellipsometry measures in the polarization state of light when it is reflected from a surface of a sample. As a result, $\Psi$ (relative amplitude change) and $\Delta$ (relative phase change) spectra can be obtained from the measurements. In this investigation, the measurements of $\Psi$ and $\Delta$ spectra have been carried out in a frequency range between 300 and 6000~cm$^{-1}$ within an angle of incidence for 60$^{\circ}$ and a resolution of 4~cm$^{-1}$ under dry air (IR-WASE: J.A. Woollam Col, Inc.). These $\Psi$ and $\Delta$ parameters are corresponding to the Fresnel reflection coefficients $R_{\rm p}$ (p-polarized: parallel to the plane of incidence) and $R_{\rm s}$ (s-polarized: perpendicular to the plane of incidence). So the complex reflectance ratio $\rho$ as the basic equation for ellipsometry can be described as
%
\begin{equation}
{\rho=\frac{R_{\rm p}}{R_{\rm s}}={\rm tan}(\Psi)~{\rm e}^{-i\Delta}}
\end{equation}
%
where tan($\Psi$) denotes the ratio of reflected amplitudes and $\Delta$ is the phase shift of p- and s- linearly polarized components due to reflection.

\begin{figure*}

\begin{center}
\includegraphics[width=0.49\hsize]{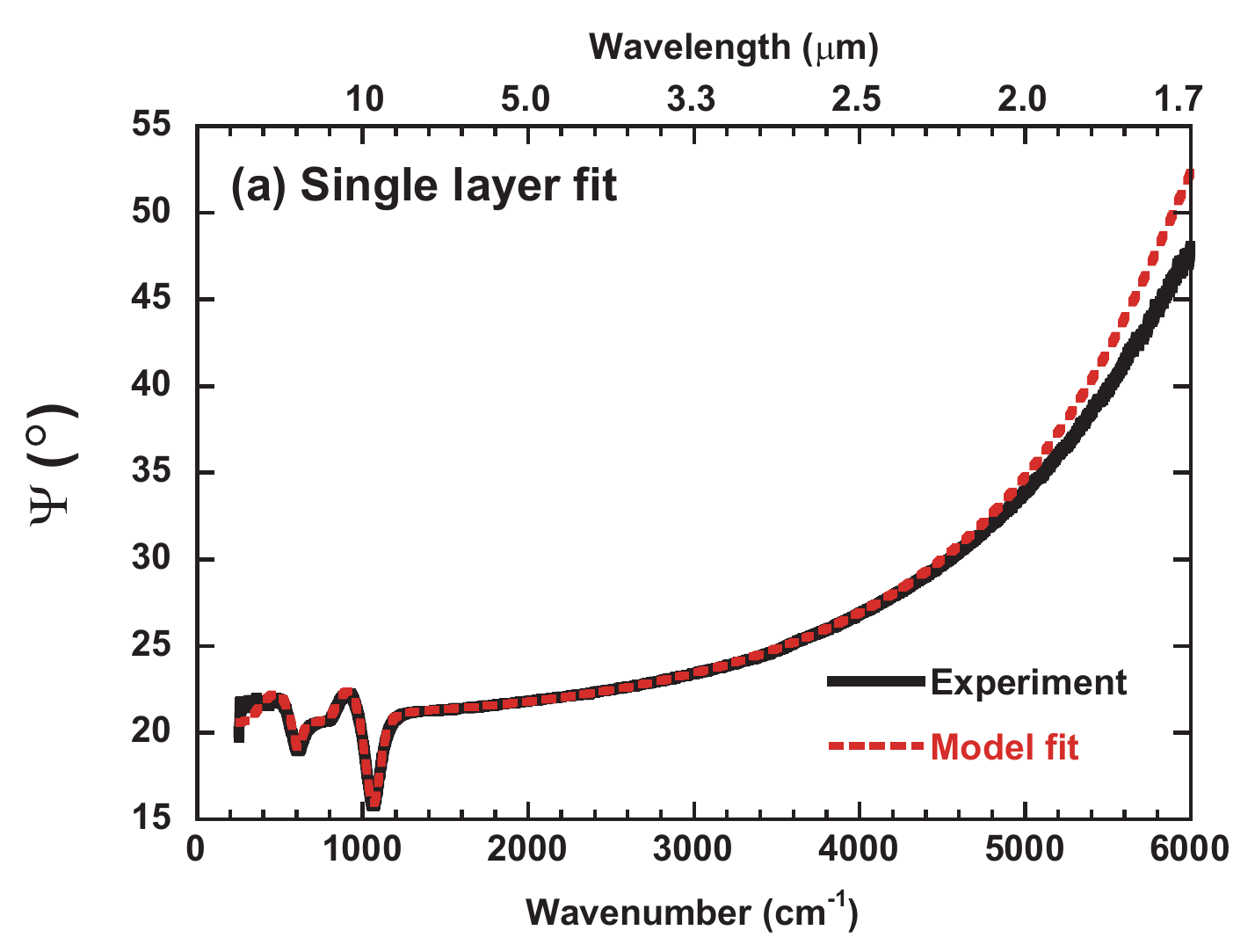}
\includegraphics[width=0.49\hsize]{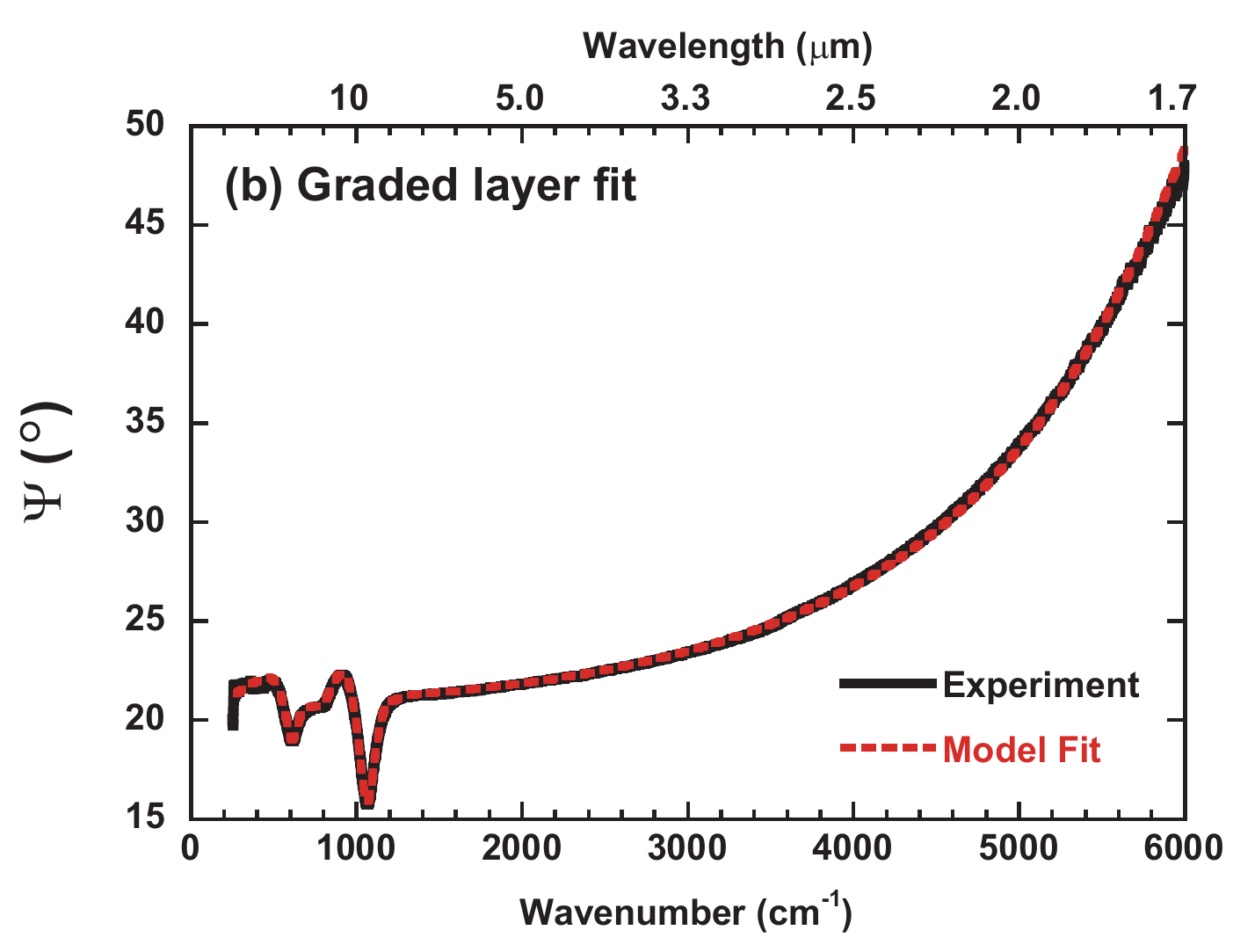}
\includegraphics[width=0.49\hsize]{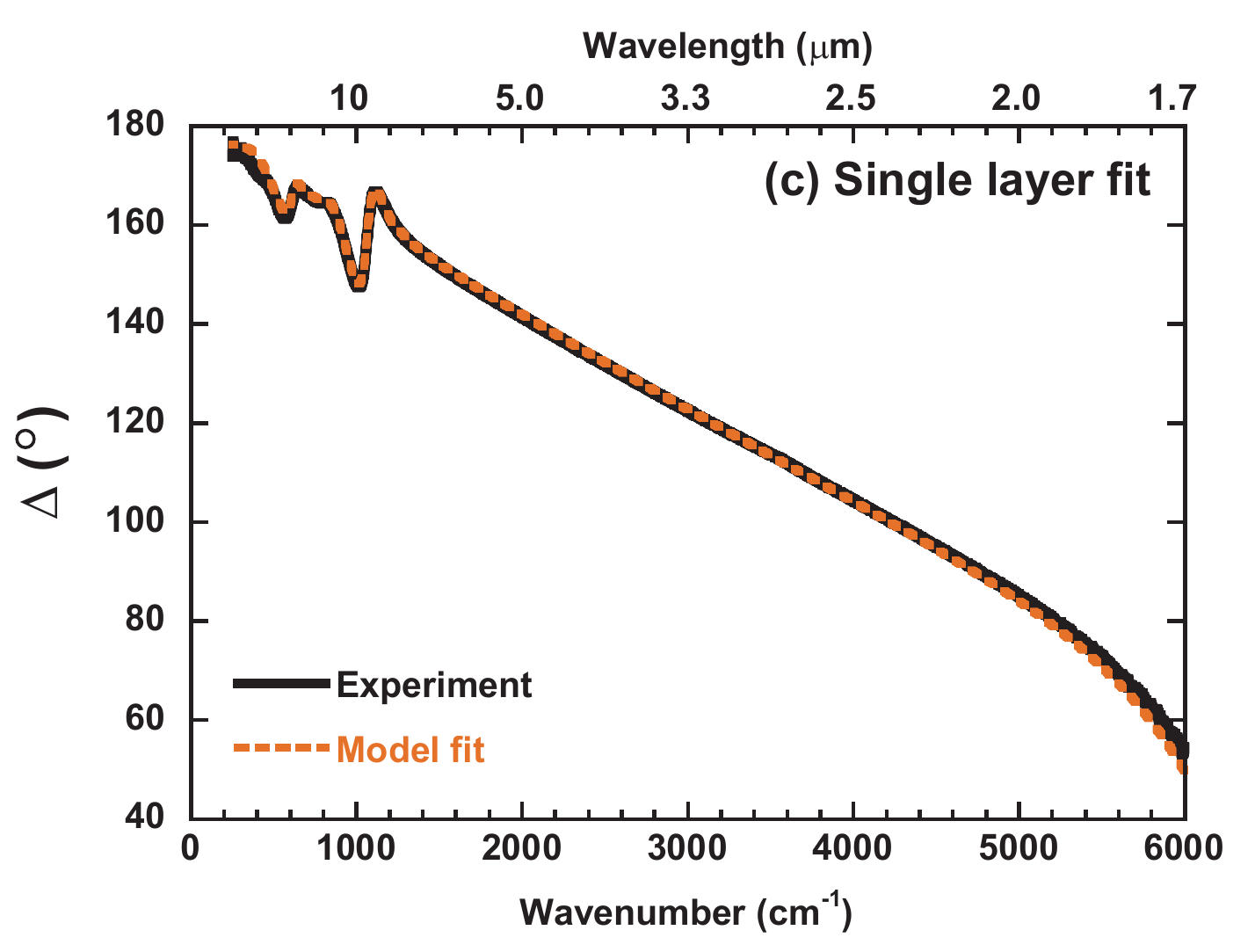}
\includegraphics[width=0.49\hsize]{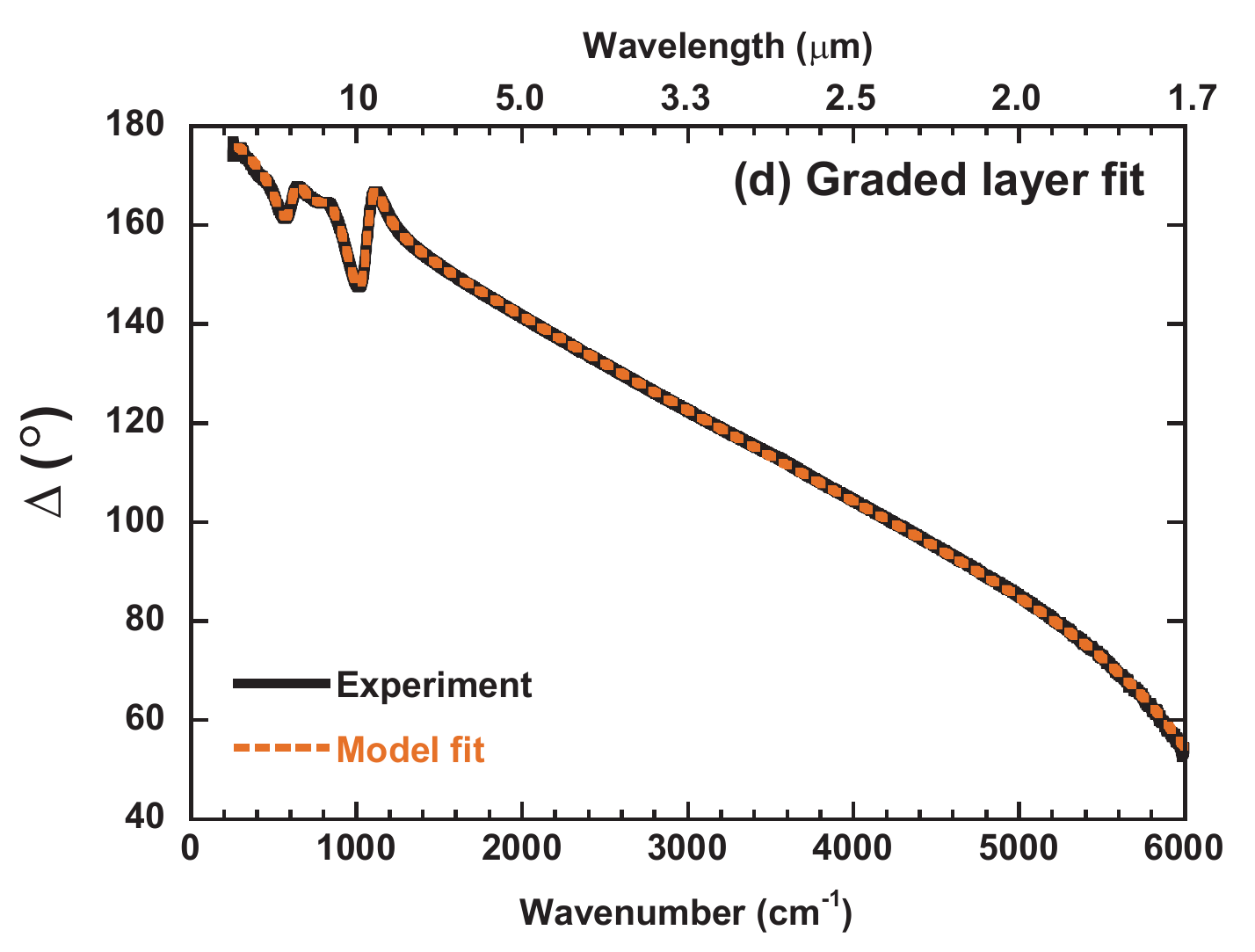}
\end{center}

\caption{Ellipsometric $\Psi$ (upper panels) and $\Delta$ spectra (bottom panels) for Si-film3 sample together with the model fit at an incident angle 60$^{\circ}$: (a) and (c) without considering the graded index effect; (b) and (d) with considering the graded behaviour. The solid likes denote the measured data whereas the dotted lines are the model fit data. }

\label{FigEllipSpec}
\end{figure*}

A dielectric function which is derived from the ellipsometry measurements for uniform and isotropic balk materials, is described as a ``pseudo-dielectric function" that can be derived from the measured values of  $\Psi$ and $\Delta$ spectra,
%
\begin{equation}
{\varepsilon=\varepsilon_1+\varepsilon_2={\rm sin^2}(\Phi_1)\left\{1+{\rm tan}^2(\Phi_1)\Biggl(\frac{1-\rho}{1+\rho} \Biggr) \right\}}
\end{equation}
%
where $\varepsilon_1$ is the real part and $\varepsilon_2$ is the imaginary part of the complex dielectric function, $\Phi_1$ is the angle of the incident radiation \citep{Fuj07}.

Because of the inhomogeneous line broadening of the amorphous samples, the Gaussian oscillator model of the ellipsometry software is applied for modeling the IR vibrational modes of the amorphous silicates. The spectral line shape of the measured $\Psi$ and $\Delta$ spectra is approximated by a Gaussian line profile which can be expressed by the Kramer-Kronig relations. The real part $\varepsilon_1$ of the complex dielectric function is related to the imaginary part $\varepsilon_2$ at all frequencies and is represented as 
%
\begin{equation}
{\varepsilon_1(\omega)=\frac{2}{\pi}\,P\int_{0}^{\infty}\frac{\Omega\,\varepsilon_2(\Omega)}{\Omega^2-\omega^2}d\Omega}
\end{equation}
%
where $P$ is the Cauchy principal part \citep{Pei91}.

The imaginary part describes each Gaussian oscillator mathematically 
%
\begin{equation}
\varepsilon_2(\Omega)=A{\rm e}^{-\frac{(\Omega-\omega_0)^2}{2\sigma^2}}
              -A{\rm e}^{-\frac{(\Omega+\omega_0)^2}{2\sigma^2}}
\end{equation}
%
where $A$ denotes the amplitude, $\omega_0$ is the centre/resonance energy, and $\sigma$ is related to the broadening $Br$ of a band by
%
\begin{equation}
{\sigma=\frac{Br}{2\sqrt{2\ln(2)}}}
\end{equation}
%
\citep{DeS06}. All three parameters, $A$, $\sigma$, and $\omega_0$, are ellipsometry model fit values which are summarized in Table~\ref{table: 3}.

In this analysis, we consider a very slight composition gradient with silicate thickness by applying the ``graded index layer model'' of the WVASE software. Such composition gradient depth might occur due to, e.g., different oxygen contents (e.g., oxidation from atmosphere) and/or surface roughness (more voids on a surface)  \citep{Rot98,Syn98}. Figure~\ref{FigEllipSpec} shows a comparison between the Gaussian model fit without and with considering the graded effect.

\begin{figure}
\begin{center}
\includegraphics[width=\hsize]{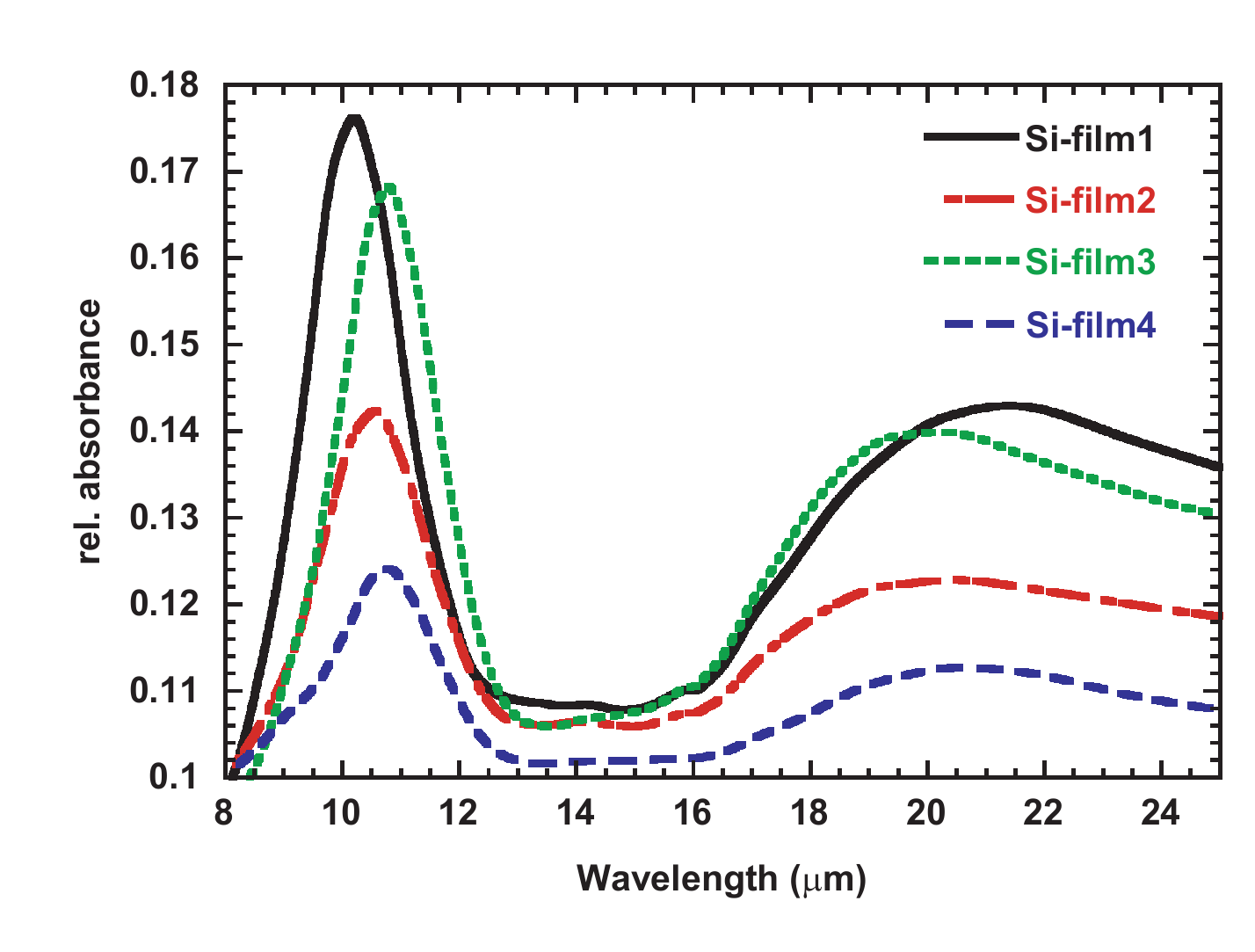}
\end{center}

\caption{Relative absorbance spectra of four amorphous silicate samples with different stoichometries. Note: the solid line corresponds to Si-film1, the long dash-dotted line  to Si-film2, the short dashed line to Si-film3), and the long dash line to Si-film4.}
\label{FigRelAbs}
\end{figure}

A clear deviation could be seen at longer ($\gtrsim20\ \mu$m) and shorter ($\lesssim2\ \mu$m) wavelength regions in Figure~\ref{FigEllipSpec} (a) and (c). Especially, the discrepancy at shorter wavelength region influences the accurate determination of the real part value of the dielectric function at large photon energies ($\varepsilon_{\infty}$). In ellipsometry measurements, the value of the high frequency dielectric constant ($\varepsilon_{\infty}$) is determined as a fit parameter. The value of $\varepsilon_{\infty}$ increases with increasing the concentration of Fe in the amorphous silicate system (Table~\ref{table: 3}). When the ``graded index layer model'' is not applied for the model fitting, the  $\varepsilon_{\infty}$ value of the Si-film3 is 3.09 which is 0.06 lower than the ``graded layer'' value. The maximum difference in the $\varepsilon_{\infty}$ value between the graded layers and single layer fitting is 0.12 which is obtained for the Si-film4. Furthermore, in the case of silicates, the high accuracy fitting at longer wavelength region is necessary so as to pin down the exact location of the Si-O bending vibration band at around 20 $\mu$m. Therefore, the graded index effect is taken into account for improving the model fit of silicate samples as shown in Figure~\ref{FigEllipSpec} (b) and (d).

There are several discrepancies between the oscillator data of Table \ref{table: 2} and \ref{table: 3}. Whereas transmittance at normal incidence measures only vibrational dipoles parallel to the layer, ellipsometry is sensitive also to those perpendicular to the layer. In our fit we assumed isotropic and smooth layers and, hence, any small anisotropy and inhomogeneity is not taken into account in the ellipsometric analysis. Nevertheless, ellipsometry delivered accurate $\epsilon_\infty$ values the use of which improved the transmittance analysis.

\begin{table}	
\centering
\caption{Si-O-Si vibrational assignments of the strongest bands.}
\label{table: 4}
\normalsize
\begin{tabular}{c c c}	
\hline
\hline
\noalign{\smallskip}
Samples      &   Asymmetric Stretching          & Bending         \\
		     &  [$\mu$m]             & [$\mu$m]        \\
\noalign{\smallskip}
\hline
\noalign{\smallskip}
Si-film1          &   10.2               & 21.4	   \\	
Si-film2          &   10.6               & 20.5    \\
Si-film3          &   10.8               & 20.2    \\
Si-film4          &   10.8               & 20.5    \\
\noalign{\smallskip}
\hline
\end{tabular}

\end{table}

\section{Experimental results}

\subsection{Absorbance behaviour}
\label{SectOptConst0}

\begin{figure*}
	\begin{center}
		\includegraphics[width=0.49\hsize]{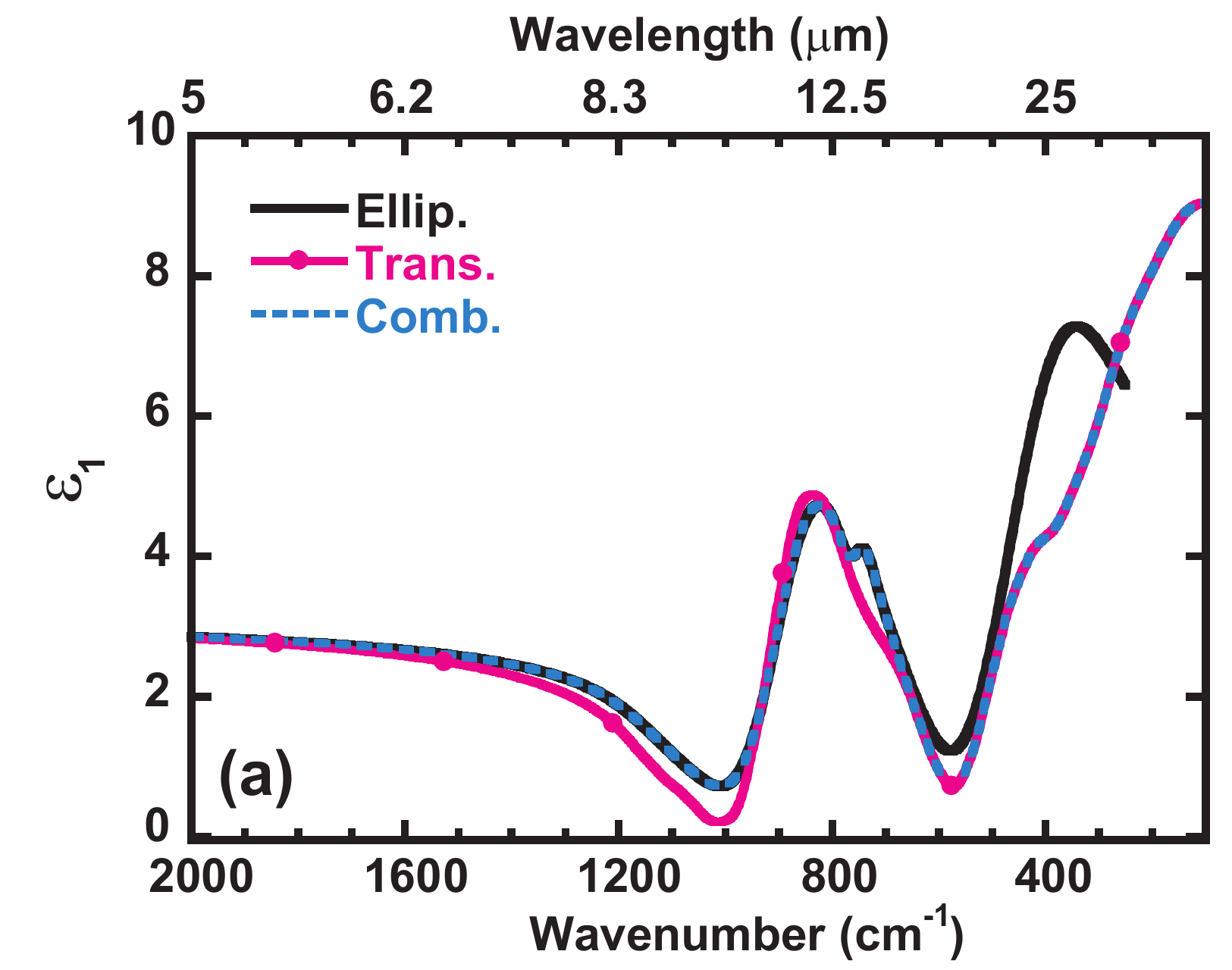}
		\includegraphics[width=0.49\hsize]{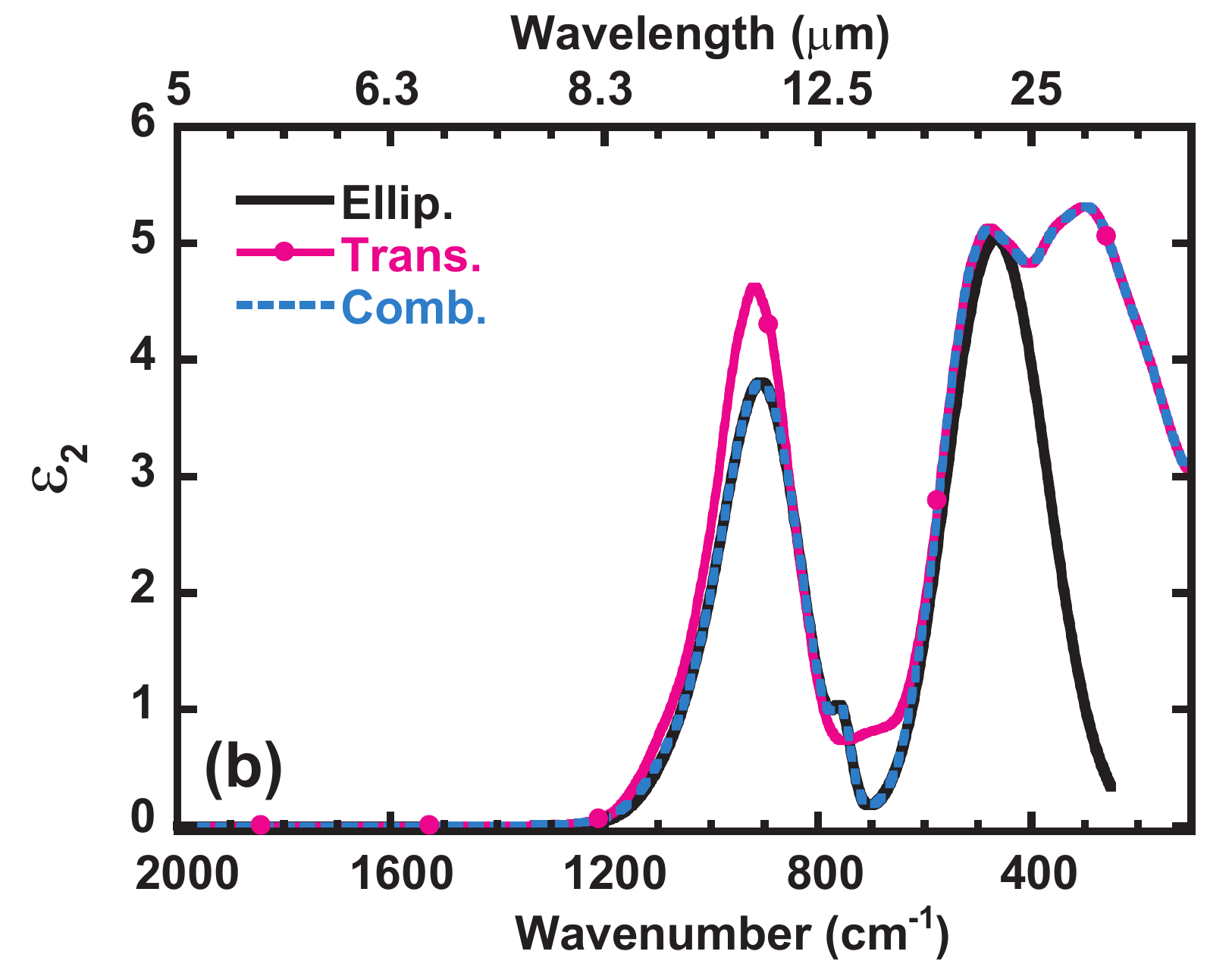}
	\end{center}
\caption{Dielectric modeling of ellipsometric and transmittance spectra as well as the best combined data of both models for Si-film3 sample: (a) The real part of the dielectric function; (b) the imaginary part of the dielectric function The solid line corresponds to the ellipsometric model, the solid line with circle marks to the transmittance line model, and the short dash line to the combination of both models.}
\label{FigDielMod}
\end{figure*}

In Figure~\ref{FigRelAbs} the measured transmittance spectra have been converted into absorbance spectra by the equation
\begin{displaymath}
 \mathrm{absorbance}=-\log( T)
\end{displaymath}
($T$ is the transmittance). Figure~\ref{FigRelAbs} shows these absorbance spectra in the wavelength between of 8 and 25~$\mu$m for four various stoichiometries of amorphous silicate samples. Table~\ref{table: 4} lists the positions of the two strongest absorbance features. They mainly include Si-O-Si stretching and bending vibrations, respectively. As obvious in Figure  \ref{FigRelAbs}, the Si-O asymmetric stretching vibration band undergoes a redshift as the Fe concentration in the amorphous silicate system increases. However, this tendency is not systematic up to Si-film4 because of the different behaviour of the two modes hat contribute to that peak, see Table~\ref{table: 2}. A weak Si-O-Si symmetric stretching vibration band around 13 -- 14~$\mu$m has been observed for Si-film1, Si-film2, and Si-film3.

\subsection{Derived Optical Constants}
\label{SectOptConst}

As mentioned in Sec.~\ref{SectFTIR} and \ref{SectEllips}, we have derived the optical constants of each amorphous silicate film by two different methods. We inspect whether these methods deliver corresponding optical constants and how we can combine the results for further astronomical simulations in this section.

A comparison of the real and imaginary parts derived from modeling the ellipsometric and transmittance spectra of Si-film3 sample is shown in Figure~\ref{FigDielMod} (a) and (b), respectively. Since we employed the value of $\varepsilon_{\infty}$ derived from the ellipsometric model for the transmittance spectral fit could be improved significantly.  Concerning the imaginary part, the peak positions of the resonance frequency around 10~$\mu$m as well as the peak around 20~$\mu$m are located at almost the same positions. However, there is an obvious difference observed around 12.5~$\mu$m: Absorption due to the Si-O symmetric vibration clearly appears as a shoulder in the $\varepsilon_2$ spectrum derived from the ellipsometric modeling, but in the transmittance fit it appears only as a plateau between two strong features. The fit parameters for this resonance position are different for the two methods, see Table \ref{table: 2} and  \ref{table: 3}. The same result has been obtained for Si-film1 and Si-film2 samples. These deviations are related to the very weak IR activity of this mode which together with the inhomogeneous line broadening  complicates the spectral analysis. The strong disparity at the longer wavelength side ($\gtrsim20\ \mu$m) is related to the limited measurement range of the IRSE which is 1.7 -- 33~$\mu$m. If we approach the limit of the measurement range ($\gtrsim20\ \mu$m), the signal-to-noise ratio decreases, see this spectral range in Figure \ref{FigEllipSpec}. In this noisy range the ellipsometric data are not used anymore for further simulations.

\begin{table*}

\caption{The sample of supergiants and their basic data.}

\begin{tabular}{lcccccclrl}
\hline
\hline
\noalign{\smallskip}
     & Spectral\tablenotemark{a} & Variable\tablenotemark{a} & Period\tablenotemark{a} & $T_\mathrm{eff}$\tablenotemark{b} & $L$\tablenotemark{b} & $\dot M_\mathrm{gas}$\tablenotemark{c} & $v_\mathrm{exp}$\tablenotemark{c} & $D$\tablenotemark{d} & $A_V$\tablenotemark{b} \\
\noalign{\smallskip}
Name & type & type & [d] & [K] & [L$_{\sun}$] & [$\rm M_{\sun}a^{-1}$] & [km\,s$^{-1}$] & [pc] & \\
\noalign{\smallskip}
\hline
\noalign{\smallskip}
$\mu$ Cep & M2Ia     & SRc & 730 & 3\,700 & $3.4\times10^5$  & $5.0\times10^{-6}$ & 20 &    870 & 2.01 \\
RW Cyg    & M3Iab    & SRc & 550 & 3\,600 & $1.45\times10^5$ & $3.2\times10^{-6}$ & 23 & 1\,320 & 4.49 \\
W Per     & M3Iab & SRc    & 485 & 3\,550 & $5.5\times10^4$  & $2.1\times10^{-6}$ & 16 & 1\,900 & 4.03 \\
RS Per    & M4Iab    & SRc & 245 & 3\,550 &  $1.4\times10^5$ & $2.0\times10^{-6}$                    &  20  & 2\,400 & 2.63 \\
\noalign{\smallskip}
\hline
\end{tabular}
\tablenotetext{a}{Taken from SIMBAD database.} \tablenotetext{b}{\citet{Lev06}.} \tablenotetext{c}{\citet{Mau11}.} \tablenotetext{d}{\citet{Jon12}.} 

\label{TabStarParm}
\end{table*}

Both derived data sets were combined at reasonable points in order to cover a broad spectral range between 2 and 200 $\mu$m for astronomical simulations. In Figure~\ref{FigDielMod}, the short dashed line exhibits the combined dielectric function. For both the real and imaginary parts, there were some points of contact where it was easy to connect these data sets without losing accuracy critically. Figure \ref{FigDielMod}a shows the real part of the Si-film3. The transmittance data were utilized at the short wavelength between 1.25 and 4.17~$\mu$m. The ellipsometric model data were integrated at 4.17~$\mu$m up to 14.3~$\mu$m because the Si-O-Si symmetric stretching vibration band more clearly appears in this region. Again the ellipsometric data were combined with the transmittance data at the point of 15.4~$\mu$m. In here, because there was no exact intersection point between those two data sets, we interpolated the ellipsometric data between 14.3 and 15.4~$\mu$m where the ellipsometric data line was descended along the transmittance line curve up to the 15.4~$\mu$m. Then the transmittance data were selected from 15.4~$\mu$m to 200~$\mu$m. As mentioned in Sect.~3.3, the $\Delta$ spectra are highly sensitive to thin films; therefore, we select the ellipsometric data in the MIR region where the Si-O-Si asymmetric and symmetric vibration bands obviously appear. A similar modification of the complex dielectric function data sets was performed for Si-film1, Si-film2, and Si-film4.

\section{Application to circumstellar dust shells}

We calculate synthetic spectra for dust enshrouded evolved stars using the newly derived optical constants for non-stoichiometric amorphous silicate materials with similar to compositions intermediate between olivine and pyroxene and compare the results with observed IR spectra from stellar sources taken with the ISO satellite. We aim to check how well one can match the observed emission properties of circumstellar dust with models for the infrared emission based on our newly determined optical constants.

\subsection{Choice of Model Stars}

Dust associated with highly evolved cool stars is freshly formed in the massive outflows that evolve as the stars climb up along the Red Giant Branch in the Hertzsprung-Russell diagram (HR) of stellar evolution towards very high luminosities. Depending on whether the oxygen to carbon abundance ratio remains $\lesssim1$ or exceeds this value, the stars form mineral dust from the abundant rock-forming elements (Mg,Si,Al,Ca,Fe) or they form soot and carbides, respectively. 

The low and intermediate mass stars (with initial masses $\lesssim8\,\rm M_{\sun}a^{-1}$) are during their dust-forming phase on their second Giant Branch, the Asymptotic Giant Branch (AGB), where they have a degenerated carbon-oxygen core and burn alternatively hydrogen and helium in two shell sources. Part of them become carbon stars (with C/O~$\gtrsim1$ at their surface) if some of the freshly produced carbon is dredged-up to the stellar surface. The massive supergiants (with initial masses $\gtrsim8\,\rm M_{\sun}a^{-1}$ and $\lesssim40\ \rm M_{\sun}a^{-1}$) are on their first Giant Branch where they burn hydrogen in a shell source on top of their helium core. Owing to their different mass and internal structures, the properties of AGB stars and supergiants are somewhat different during their final evolution. The most important differences in our context are that supergiants (i) retain their initial abundances of refractory dust forming elements during their whole evolution until they explode as a supernova, and (ii) they have much higher luminosities and slightly higher surface temperatures than AGB stars such that they do not cross one of the instability strips in the HR-diagram where stars become large amplitude pulsators. These differences makes supergiants better suited test beds for investigations of dust condensation in stars than AGB stars.

First, the critical factor ruling the chemistry in the outflow, the C/O abundance ratio, does not increase by mixing substantial amounts of carbon  from the core region to the surface as in AGB stars, but instead it slightly decreases by mixing CNO-cycle equilibrated material in which most C is converted to $^{14}$N to the surface. This guarantees that there is always a sufficient excess of oxygen over carbon that all the refractory rock-forming elements can be oxidized. This leaves no doubt on the chemical nature and composition of the dust mixture which could be formed.

Second, though the supergiants are variable to some extent as all stars in the right upper part of the HR-diagram, they do not show the high-amplitude visual magnitude variations characteristic of the pulsationally unstable stars found on or close to the AGB \citep{Arr15}. For this reason, the hydrodynamic structure of the outflow from supergiants is expected to be much simpler than for their lower-mass relatives on the AGB because there are no strong shock waves running through the atmosphere and the inner part of the dust envelope as in the case of Miras or semi-regular variables. A simple steady outflow model represents the density structure of the dust shell sufficiently accurate for our purpose. 

We restrict our considerations to objects with optically thin dust shells. This avoids that radiative transfer effects and scattering are important for the formation of the IR emission spectrum which complicates the interpretation of the spectra. 

Dust in the environment of supergiants has been studied by \citet{Spe00} and \citet{Ver09} for a large sample of stars for which ISO spectra are available. We select from these studies a small number of objects which satisfy the following requirements:
\begin{enumerate}

\item The dust shells are obviously optically thin in the IR region. For this, we consider only objects of the category 2.SEc as defined by \citet{Kra02}.  

\item The spectra show marked, but not very strong, emission  bands around 10 $\mu$m and 18 $\mu$m that are typical for amorphous silicate dust. This serves to ensure that the optical thickness in the bands is $\ll1$. 

\item The spectra show no obvious indication of the presence of crystalline silicates to avoid a significant contribution of such material to the bands around 10 $\mu$m and 18~$\mu$m. 

\item The spectra show low noise in the spectral region $\lambda\lesssim25\ \mu$m. 

\end{enumerate}
The objects that fulfil these requirements such that they are suited for our purpose belong to the objects of the category 2.SEc as defined by Kramer et al. (2002) which is essentially defined as satisfying the properties of items 1 to 3. Our set of objects selected from the database of Sloan et al (2003) is shown in Table \ref{TabStarParm} together with some basic parameters of the star and the outflow. Completely reduced ISO-spectra for these four stars, as used in \citet{Ver09}, were kindly provided to us by S. Hony.

We fit the MIR spectral data of these four test objects with calculated spectra from a model for an optically thin circumstellar dust shell. The properties of the dust shell including the dust mixture are determined by an optimization procedure where we vary the parameters of the problem such as to obtain an as good as possible fit between observed and synthetic spectrum.  

\subsection{Model Assumptions}

We do not model in this study the formation of dust in the stellar outflow and the acceleration of the outflow by radiation pressure on the dust particles and subsequent momentum transfer from dust to gas by frictional coupling. Instead, we follow the practice of most other laboratory studies which compare their results with observed stellar spectra and assume a spherically symmetric dust density distribution around the star and instantaneous condensation of dust once the dust temperature drops below a given condensation temperature.  For optically thin dust shells such a simple model is also not as unrealistic as it looks at a first glance because the details of the spatial distribution of the dust are of minor importance (see the Appendix \ref{AppRad}).

The main purpose is to check how well the amorphous silicates for which we determined optical constants can reproduce the observed MIR emission spectrum of our test stars. The amorphous silicate dust generally dominates the IR-emission from circumstellar dust in the 9 to 20 $\mu$m spectral region where also the diagnostically important resonances from stretching vibrations of the Si-O bond ($\sim10\ \mu$m) and O-Si-O bending ($\sim18\ \mu$m) vibrations are located. For fitting the synthetic model we therefore use the spectral region between 9 $\mu$m and 25 $\mu$m. Further we consider in the fitting procedure the spectral region between 3 $\mu$m and 4.5 $\mu$m because (i) in this range the shape of the observed spectrum shows no suspiciously strong molecular absorption bands that render it difficult to find the level of stellar continuum emission, and (ii) the absorption of silicate dust in this range is very low such that one observes the un-attenuated stellar radiation. This NIR spectral range helps to fix the baseline of zero dust emission.  

\begin{figure}
\includegraphics[width=\hsize]{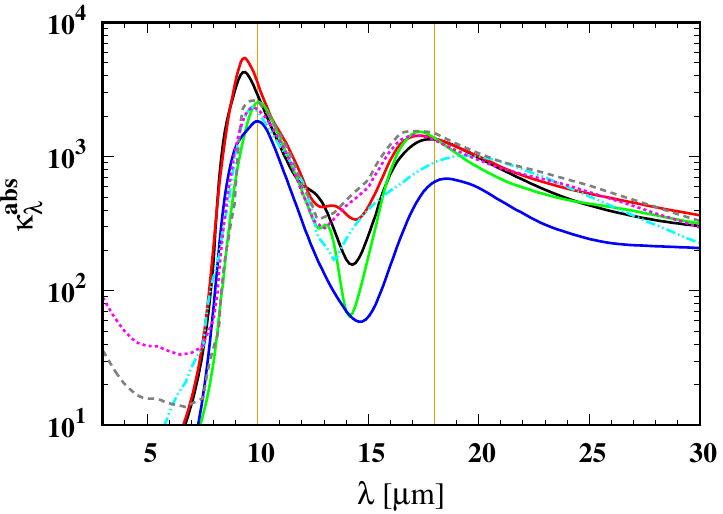}

\caption{Mass absorption coefficient $\kappa_{\lambda}^{\rm abs}$ for an ensemble of spherical silicate particles with composition Si-film1 (black line), Si-film2 (red line), Si-film3 (green line), and Si-film4 (blue line ) (all shown as full lines). For comparison the $x=.5$ data from \citep{Dor95} (dotted lilac line) and the $x=0.7$ (dashed gray line) and $x=1.0$ (dash-dotted cyan line) data from \citet{Jae03} are shown. 
}

\label{FigQabsSil}
\end{figure}

Figure \ref{FigQabsSil} shows the mass-absorption coefficients calculated for the four different amorphous silicate materials discussed in Sect.~\ref{SectOptConst0}. For the averaging over particle sizes a MRN distribution between a minimum diameter of 10 nm and a maximum of 500 nm in ten bins is used for all species. In applications to circumstellar dust shells it usually turns out that the assumption of spherical grains is not suited to reproduce observed band profiles. These are much better reproduced by using, e.g., a continuous distribution of ellipsoids (CDE) \citep{Boh83} which is therefore used in our calculations for all species, except for iron, for which we use Mie-theory. We compare the absorption coefficient calculated for or amorphous silicates with absorption coefficients using data for optical constants of amorphous silicates from \citet{Dor95} and \citet{Jae03} which are synthesized by methods different from ours.

For calculating dust temperatures one needs to know the optical properties also in the range $\lambda$  $<$ 2 $\mu$m where we have no measured data for our silicate materials. We use in the 0.2 $\leq$ $\lambda$ $<$ 4 $\mu$m wavelength range for the purpose of calculating dust temperature the \textit{x} = 0.5 data from \citet{Dor95}. The cut at $\lambda$ = 4 $\mu$m is chosen such that it is located in a wavelength range where the silicate opacity is very low such that it contributes neither to dust heating nor to emission from dust. This choice is somewhat arbitrary, but it influences only the calculated radius of the inner edge of the dust shell for given condensation temperature, but not the infrared model spectrum at given condensation temperature and dust mass.

\begin{figure*}

\centerline{
\includegraphics[width=.8\hsize]{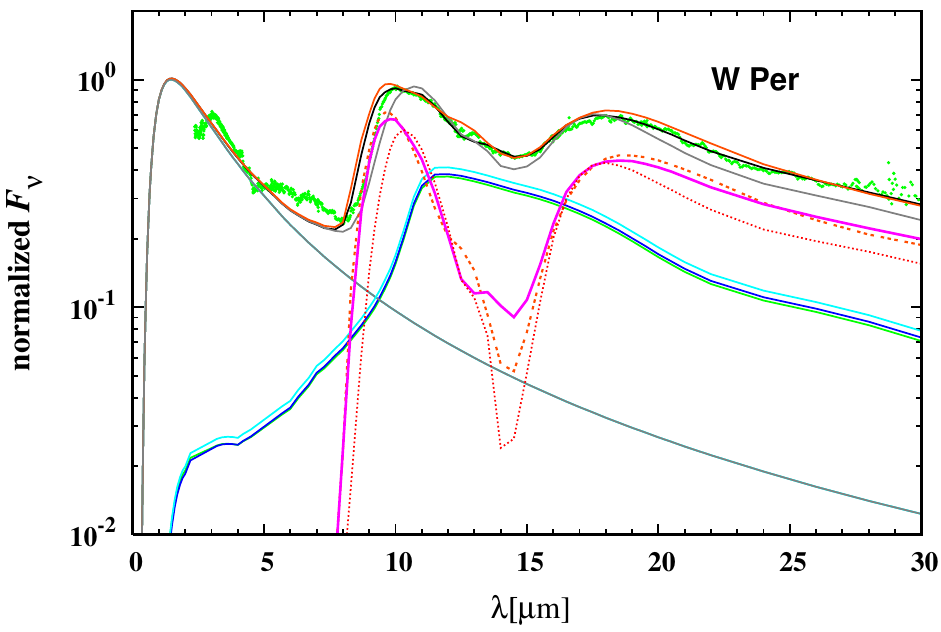}
}

\caption{%
Normalized fits to the spectrum of W Per using the optical data of Si-film1, Si-film2, and Si-film3 for the amorphous silicate component. Green dots correspond the ISO data. The orange, black, and grey curves correspond the optimized synthetic spectrum using Si-film1, Si-film2, and Si-film3 silicate material, respectively. The full magenta line and the dashed and the dotted red lines show the separate contributions of the silicate materials in the same order and the cyan, blue, and green lines the contribution of the corundum dust to the total emission, also in the same order. The light grey line shows the stellar black-body spectrum.
}

\label{FigWPer-3-spec}
\end{figure*}

In principle it would be desirable to use in the fitting process only the amorphous silicate dust material studied in this paper since we are interested in the first place whether it is possible to get reasonable fits already with these materials alone. However, one finds in most oxygen-rich objects also a non-negligible contribution to emission from an additional dust component which is identified with amorphous corundum \citep{Ona89,Ona89b,LoM90, Jon14}. Unfortunately the emission from corundum contributes to the spectral range between 10 and 15 $\mu$m such that it interferes with the silicate dust emission. In particular it fills up the deep absorption minimum between the two silicate features. For this reason we include amorphous corundum as a dust component in the modeling. 

Optical constants for the corundum for $\lambda\ge7.8\ \mu$m are taken from \citet{Beg97} (the constants for porous amorphous aluminium oxide). For shorter wavelengths they are augmented by data from \citet{Pal85}. 

Interferometric observations have shown that the corundum dust seems to be formed much closer to the star than the silicate dust \citep[e.g. ][ and references therein]{Dan94,Kar13,Wit15}, as it could be expected according to its much higher thermal stability \citep{Sal77}. The grains then either may be heterogeneously composed of corundum grains condensing earlier than silicates in a stellar outflow which later become overgrown by silicate material as temperature drops in the outflow, or they are the result of agglomeration of corundum grains and silicate grains. For modeling the opacity we have to include corundum grains either as  a separate dust component, or as inclusions within silicate grains, or in both variants. Practically we found to include the corundum only in form of a separate dust component to result in slightly better fits than to include also silicate particles with corundum inclusions. The difference is, however, only marginal such that one cannot clearly discriminate between the possible cases on the basis of spectral modeling alone. We consider in our modeling corundum and amorphous silicate as separate dust components. 

Further, metallic iron particles seem to be present as inclusions in the silicate grains. They are required to explain the NIR opacity of real circumstellar dust \citep{Oss92} and the elemental composition of presolar silicate grains (see Sect.~\ref{SectPresolar}). The contribution of such inclusions to the opacity of the composed particles is low in the diagnostically important spectral region at wavelengths longer than 9 $\mu$m and they cause in that region only an almost constant upward shift of dust opacity that cannot be discriminated from the effect of a slight increase of silicate dust abundance. The main effect of such iron inclusions is its strong contribution to the optical to NIR dust opacity which rules the dust temperature. But this effect of the iron inclusions becomes really important only for an iron-free silicate matrix. If the silicate material already contains a significant iron content, the influence of possible iron metal inclusions becomes moderate. We found that one cannot clearly discriminate by the quality of fit between models with and without iron grains because of a degeneracy between content of iron inclusions an amount of dust present. We therefore neglect iron inclusions.

\begin{figure}[t]

\includegraphics[width=\hsize]{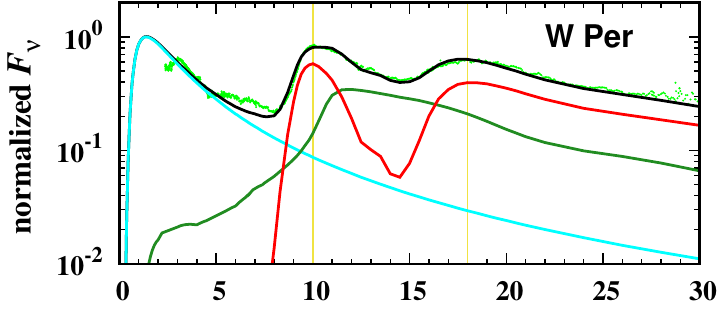}

\includegraphics[width=\hsize]{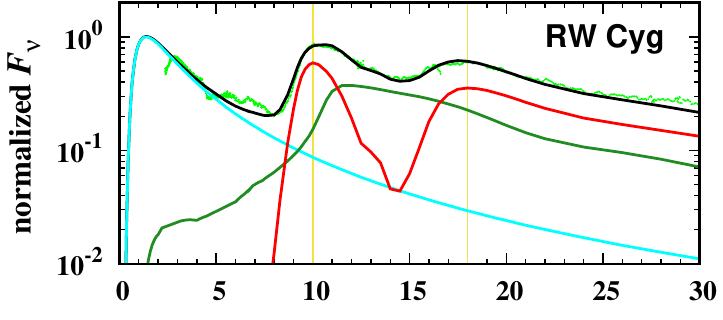}

\includegraphics[width=\hsize]{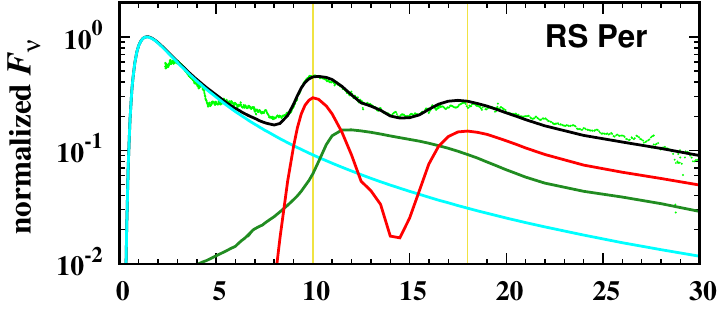}

\includegraphics[width=\hsize]{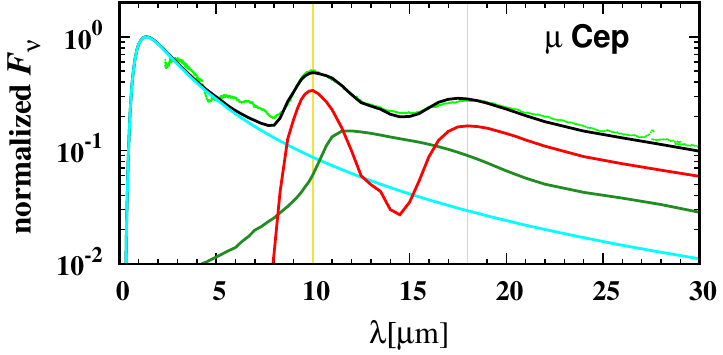}

\caption{ISO spectrum of supergiants fitted with a synthetic spectrum of a star with circumstellar dust shell. The small dots show the ISO-sws spectrum, the black line the optimum spectral fit, the cyan line the stellar black body spectrum. Also shown are the individual contributions of the dust species to the spectrum: Silicate dust (red line) and amorphous corundum (green line). For the silicate material, in all cases the optical data of Si-film2 and Si-film3 are interpolated with fractions varying from star to star, see text or the value of $g$ in Table \ref{TabModFit} for this.}

\label{FigSupGiantSpec}
\end{figure}

Another possibility is that iron forms as a separate dust component. This follows from considerations on thermal stability \citep[e.g.][]{Sal77,Gai99} and has been found to be a likely contributor to opacity in circumstellar dust shell in the $\lambda<8\ \mu$m spectral region by \citet{Kem02} and \citet{Ver09}. This case differs from the case of iron inclusions in so far, as inclusions are thermally coupled to the silicate matrix while separate iron particles have their own temperature distribution. In effect, this results in significant differences in the calculated model spectra compared to the case of iron inclusions. We also consider iron grains in our modeling. The optical constants are taken from \citet{Ord88}.

\begin{figure}[t]

\includegraphics[width=\hsize]{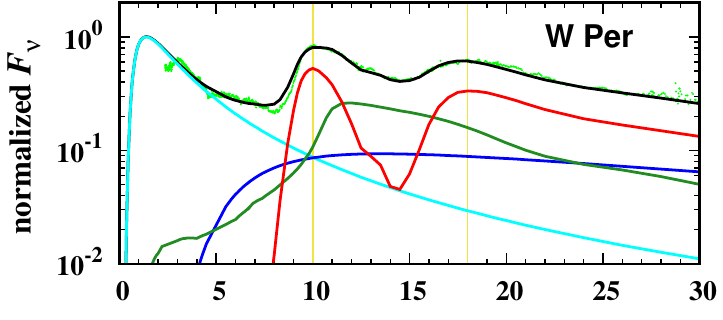}

\includegraphics[width=\hsize]{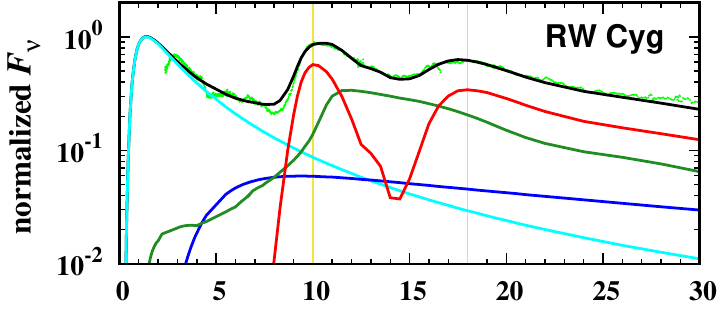}

\includegraphics[width=\hsize]{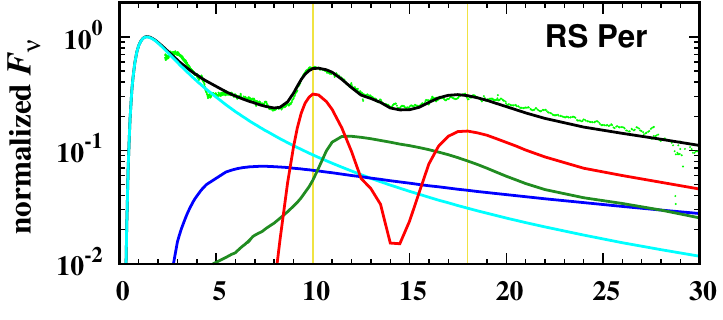}

\includegraphics[width=\hsize]{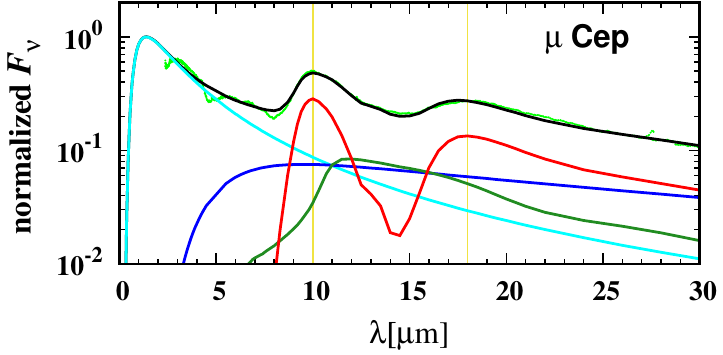}

\caption{Same models as in Figure~\ref{FigSupGiantSpec} with metallic iron micrograins (blue line) as additional dust component.. }

\label{FigSupGiantSpec3}
\end{figure}

\subsection{Individual Fits}

We fit the MIR spectral data of the four test objects with calculated spectra from circumstellar dust shell models (also called synthetic spectra) by an optimization procedure described in Appendix \ref{AppRad}. 

The stellar luminosities and effective temperatures of the objects are taken from the literature; the corresponding values and references are given in Table~\ref{TabStarParm}. No dereddening was finally performed for the ISO spectra because it turned out that the corresponding correction is very small in the wavelength range of interest for all our objects. 

The dust temperature in the models drops below 100 K at a distance of about $10^3$ stellar radii. Regions farther out do not contribute significantly to the infrared spectral region $\lambda<30$ $\mu$m used in our comparison of observed with synthetic model spectra. The outer radius of the dust shell therefore cannot be pinned down by our model optimization; it was set to a fixed value of $10^4$ stellar radii in all cases.

\subsubsection{Fits with two Dust Components}

In the following models two dust species, amorphous silicate particles and corundum particles, are considered as opacity sources.
 
\paragraph{W Per.}
A fit to \objectname{W Per} was tried first using the optical constants of Si-film1, Si-film2 and Si-film3 for the silicate component. The results are shown in Figure~\ref{FigWPer-3-spec}. The iron-free Si-film1 can be excluded because the resonance at $\sim10\ \mu$m does not fit to the data. The optical data of Si-film2 result in a good fit for the resonance at $\sim18\ \mu$m and a reasonable fit to the peak and the long-wavelength flank of the $\sim10\ \mu$m resonance. The short-wavelength flank is too extended, however. The dust model Si-film3 gives a low quality fit, but it deviates from the ISO data just in the opposite direction than the Si-film2 model. This means that an amorphous silicate of the kind studied in this paper with an intermediate iron content between Si-film2 and Si-film3 would fit the observed spectral energy distribution.

A linear interpolation of the data for Si-film2 and Si-film3 using a fraction of 0.6 of Si-film2 and a fraction of 0.4 of Si-film3 results in an rather good fit between the ISO data and the corresponding synthetic spectrum. The corresponding model is shown in Figure~\ref{FigSupGiantSpec}. The correspondence between observed spectrum and optimized synthetic spectrum appears almost perfect, except in the range between about 6 \dots\ 8 $\mu$m. In particular the two silicate bands are well reproduced by the model. 

With respect to the missing flux shortward of 8 $\mu$m see later.

\paragraph{RW Cyg.}
Also in the case of \objectname{RW Cyg} the observed stellar spectrum is embraced by the optimized models using the data of Si-film2 or Si-film3. A model using a weighted mean of the opacity data with a fraction of 0.4 from Si-film2 and 0.6 from Si-film3 results in a model which well fits the observational data. This model is shown in  Figure~\ref{FigSupGiantSpec}.

\paragraph{RS Per.}
Though, again, the observed stellar spectrum is embraced by the optimized models using the data of Si-film2 or Si-film3, it is not possible to obtain a simultaneous close fit of the 10 $\mu$m and the 18 $\mu$m silicate bands for \objectname{RS Per}. The 10 $\mu$m band is fitted reasonably well with a weighted mean of the opacity data using a fraction of 0.4 from Si-film2 and 0.6 from Si-film3. The 18 $\mu$m band is best fitted with Si-film3 alone. A compromise is to a use a weighted mean of the opacity data using a fraction of 0.7 from Si-film3 and 0.3 from Si-film2.  The resulting model spectrum is shown in  Figure~\ref{FigSupGiantSpec}. The peak position of the 18 $\mu$m band in the model appears at a position slightly shortward of the peak in the observed spectrum, and there is a noticeable discrepancy in the model flux at wavelengths longward of 20 $\mu$m. 

\paragraph{$\mu$ Cep.}
A similar situation is encountered for \objectname[mu Cep]{$\mu$ Cep}. A compromise is here to a use a weighted mean of the opacity data using a fraction of 0.8 from Si-film2 and 0.2 from Si-film3.  The resulting model spectrum is shown in  Figure~\ref{FigSupGiantSpec}. Again, the peak position of the 18 $\mu$m band in the model appears at a position slightly shortward of the peak in the observations, and also here there is a noticeable discrepancy in the model flux at wavelengths longward of 20 $\mu$m. 

\subsubsection{Fits with three Dust Components}

In the following models metallic iron as a separate dust component is added to the amorphous silicate and corundum particles in order to provide opacity in the $\lambda<8\ \mu$m spectral region where the silicate and corundum dust components are transparent. For the optimization also the wavelength range between 5.5  and 7.5 $\mu$m is included in the calculation of $\chi^2$ because iron dust is the sole source of the emission from the dust shell in this region. 
 
\paragraph{W Per.} The addition of a third dust component changes the relative abundances of amorphous silicate and corundum and, additionally, the iron content of the amorphous silicate dust required to obtain a good fit. An interpolated dielectric function using a fraction of 0.5 for Si-film2 and Si-film3 results in a close fit between observed and synthetic spectrum.  The resulting model spectrum is shown in  Figure~\ref{FigSupGiantSpec3}. Now the flux deficit in the 5 \dots\ 8 $\mu$m spectral region of the two-component models seen in Figure~\ref{FigSupGiantSpec} has completely disappeared. The remaining slight flux excesses of the synthetic spectrum over the observed spectrum at around 4.8, 6.8, and 8 $\mu$m are obviously due to molecular absorption bands of CO, H$_2$O, and SiO, respectively. The stretching and bending modes at 10 and 18 $\mu$m of the amorphous silicate are well reproduced by the model spectrum. 

\paragraph{RW Cyg.} In this case a linear interpolation of the optical constants of Si-film2 and Si-film3 with weights 0.7 and 0.3, respectively, is required for an optimum fit.  The resulting model spectrum is shown in  Figure~\ref{FigSupGiantSpec3}. The correspondence between observed spectrum and model spectrum is very close.

\begin{figure}

\includegraphics[width=\hsize]{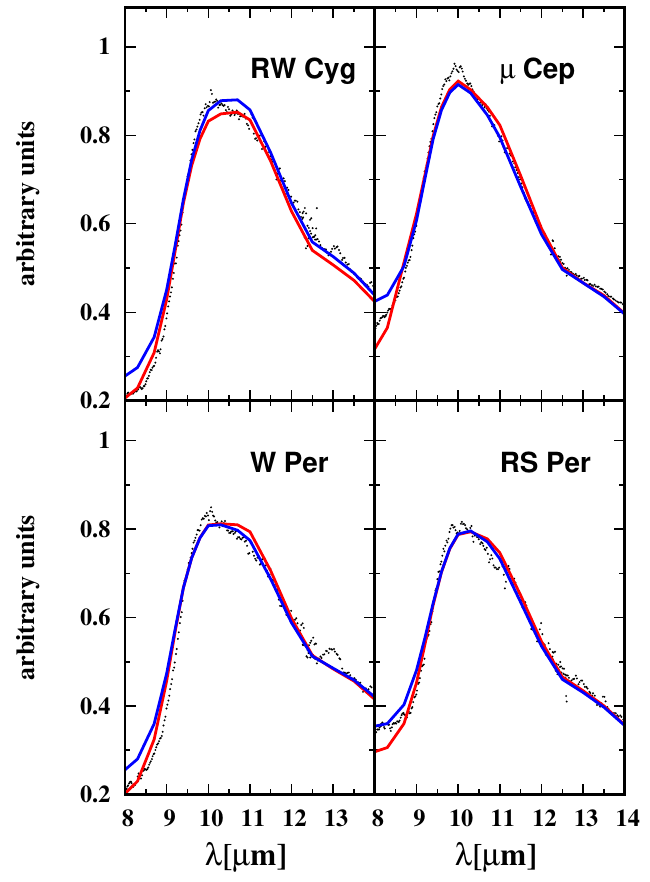}

\caption{Fit of the 10 $\mu$m silicate band. Dots are ISO spectra. The blue lines corresponds to the models with three dust components, the red lines corresponds to models with two dust components.}
 
\label{Fig10mum}

\end{figure}

\paragraph{RS Per.} In this case a linear interpolation of the optical constants of Si-film2 and Si-film3 with weights 0.2 and 0.8, respectively, is required for an optimum fit.  The resulting model spectrum is shown in  Figure~\ref{FigSupGiantSpec3}. Again, the correspondence between observed spectrum an model spectrum is good in the range between about 9 and 20 $\mu$m. At longer wavelengths the model flux is slightly below the stellar flux.

\paragraph{$\mu$ Cep.} In this case a linear interpolation of the optical constants of Si-film2 and Si-film3 with weights 0.6 and 0.4, respectively, is required for an optimum fit.  The resulting model spectrum is shown in  Figure~\ref{FigSupGiantSpec3}. The correspondence between observed spectrum and model spectrum now is almost perfect, also in the range $\lambda>20\ \mu$m.

We also tried if it is possible to reproduce the observed spectra with omitting corundum but retaining iron to fill up the silicate absorption trough between the two silicate bands. This combination gave no good fit because one can fit either the wavelength range between the two silicate features or the wavelength range between 5.5 and $\lambda$=8 $\mu$m, but not both simultaneously.

\subsubsection{The 10 $\mu$m Silicate Band}
The 10 $\mu$m band often attracts particular attention in discussions of the quality of the fit between radiative transfer models and observed spectra, so we look at this feature in more detail. 

Figure \ref{Fig10mum} shows details of the fits between observed and synthetic spectra of our comparison stars for the wavelength range around the 10 $\mu$m band corresponding to the Si-O stretching vibrations. The models using the optical data of the set of amorphous silicates synthesized and investigated by us result in fits that are able to reproduce the width, peak position, and general shape of this silicate band with considerable accuracy, if they are combined with the dust materials corundum and iron. They also reproduce the change of slope of the long wavelength flank at $\sim12$ $\mu$m which seems to be present in most supergiants \citep{Spe00}. In our model this particular detail results from a the weak Si-O-Si symmetric stretching band in the amorphous silicate absorption at this region and the broad maximum of the corundum absorption. This weak silicate feature is missing in other data sets for silicate dust absorption because the corresponding vibration mode cannot be easily separated from the background because it is very weak in absorption strength compared to the asymmetric band (see Sect.~\ref{SectEllips} and \ref{SectOptConst}).

\begin{table}

\caption{Optimized dust shell parameters.}

\begin{tabular}{@{}lrrrrrrrr@{}}
\hline
\hline
\noalign{\smallskip}
 & \multicolumn{2}{c}{Silicate} & \multicolumn{2}{c}{Corundum} & \multicolumn{2}{c}{Iron} & \\[.2cm]
Object & \multicolumn{1}{c}{$M_\mathrm{du}$} & \multicolumn{1}{c}{$T_\mathrm{i}$} & \multicolumn{1}{c}{$M_\mathrm{du}$} & \multicolumn{1}{c}{$T_\mathrm{i}$} & \multicolumn{1}{c}{$M_\mathrm{du}$} & \multicolumn{1}{c}{$T_\mathrm{i}$} &
\multicolumn{1}{c}{$\tau_{10\mu}$} & $g$ \\
\noalign{\smallskip}
\hline
\noalign{\smallskip}
W Per     & 33.1 &    554 & 16.4 & 1\,400 &     &     & .061 & .6 \\
          & 25.0 &    623 & 12.4 & 1\,400 & 53.2 & 379 & .050 & .5 \\
RW Cyg    & 23.9 &    709 & 17.7 & 1\,400 &     &     & .058 & .7 \\
          & 21.7 &    753 & 16.1 & 1\,400 & 21.7 & 485 & .053 & .4 \\
RS Per    &  3.6 &    917 &  3.8 & 1\,400 &     &     & .027 & .7 \\
          &  3.1 & 1\,087 &  2.9 & 1\,400 & 8.6 & 579 & .028 & .2 \\
$\mu$ Cep & 10.0 &    854 &  7.0 & 1\,400 &     &     & .030 & .8 \\
          &  7.1 &    959 &  4.0 & 1\,400 & 28.2 & 484 & .022 & .6 \\
\noalign{\smallskip}
\hline
\end{tabular}

\tablecomments{$M_\mathrm{du}$ is mass of the dust species in units $10^{-5}\,\rm M_{\sun}$, $T_\mathrm{i}$ is temperature of the dust species at the inner boundary (in K) , $\tau_{10\mu}$ means optical depth of the dust shell at 10~$\mu$m, and $g$ and $1-g$ are the weighting fractions of the optical constants for Si-film2 and Si-film3, respectively.}

\label{TabModFit}
\end{table}

The fits are not perfect, however, and the observed spectra show more structure than the calculated spectra which may be associated with the presence of some minor additional dust components. It is not necessary, however, to include other minerals than the components used by us to explain the main structure of the observed 10 $\mu$m band. In particular it is not necessary to invoke substantial amounts of Ca-Al-silicates (melilite) to fit the observed shape of the profile as in \citet{Ver09}. Such a compound is unlikely to exist as wide-spread dust component because otherwise it should be found in meteorites as presolar dust grains like other aluminum oxide particles, which is not the case.\footnote{
This argument would fail, however, if most of the dust produced by massive supergiants is destroyed if it is overrun by the final supernova blast wave.}
The over-all fit is, however, substantially improved if an iron dust component is included, as was already found in \citet{Ver09}.  

The quality of the fit of the 10 $\mu$m feature is nearly the same for the models with or without iron dust particles. The properties of the model with respect to dust content or iron content of the silicate material are slightly changed by including or omitting an iron dust component (cf. Table~\ref{TabModFit}), but in both cases a parameter combination can be found that reproduces the 10 $\mu$m feature. The presence of iron dust is inconsequential in this respect. 

\begin{figure}

\includegraphics[width=\hsize]{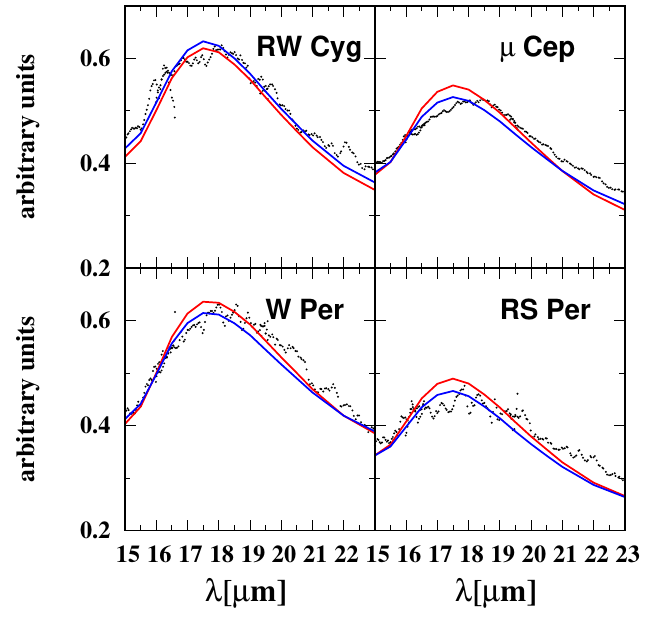}

\caption{Fit of the 18 $\mu$m silicate band. Dots are ISO spectra. The blue lines corresponds to the models with three dust components, the red lines corresponds to models with two dust components.}
 
\label{Fig18mum}

\end{figure}

\subsubsection{The 18 $\mu$m Silicate Band}

Figure \ref{Fig18mum} shows details of the fits between observed and synthetic spectra of our comparison stars for the wavelength range around the 18 $\mu$m band corresponding to the Si-O stretching vibrations. The quality of the data for the observed flux is not as good for the feature at 10 $\mu$m, but the band profiles are clearly marked and this second important emission band of silicate minerals is also reasonably fitted by the three-component mixture of our silicate materials augmented with corundum and iron, though slightly worse than the 10 $\mu$m band.    

One problem, however, is that the model flux in the spectral range $\lambda>20\,\mu$m is slightly but systematically below that of the stellar spectrum (cf. Figs. \ref{FigSupGiantSpec} and \ref{FigSupGiantSpec3}) if most weight is given to the peaks by the optimizing procedure. One reason for this deficit may be that a value of $p=2$ as assumed in our models for the density variation $\varrho\propto r^{-p}$ is unrealistic because of, e.g., variable mass-loss rate or outflow velocity. Including $p$ as an additional variable in the optimisation procedure, however, returns a value of $p\approx2.02$ and only slightly improves the situation. The flux deficit seems to be related to a deficit of the extinction coefficient of the dust mixture at long wavelength for at least one of the three dust components.

\subsubsection{Discussion}

From these results one concludes, that it is possible to explain the silicate dust emission of supergiants completely by the presence of some kind of amorphous silicate material with a non-standard stoichiometry corresponding to a composition between olivine and pyroxene. One only has to add corundum and iron dust particles in order to supply the required opacity in the range between the two silicate bands and for $\lambda<8\ \mu$m to account for the observed emission from the dust shell in these regions which cannot result from the silicate dust. Experiments with additional dust components (e.g. spinel, iron oxide) resulted in low abundances that were assigned to such additional components by the optimization algorithm.

The formation of a non-stoichiometric amorphous silicate material can be expected under the conditions of an outflow from a supergiant. Because of the high stellar atmospheric temperature the silicate dust condensation commences at considerable distances from the star when equilibrium temperatures of the dust drops sufficiently low for the dust becoming stable against vaporisation. In fact, the temperature of the inner edge of the dust shell, $T_\mathrm{i}$, in our radiative transfer model is substantially lower than an estimated vaporisation temperature of 1\,100 K, except for one case (RS Per, the model with three dust components, see Table \ref{TabModFit}) where it is close to the limit. This means that the dust is formed under conditions that resemble to some extent the vapor deposition on a cold substrate where growth species attached to the surface almost immediately become immobile at the surface of the condensed phase. 

For the kind of dust shell models calculated by us it remains open, however, if a low temperature at the inner edge, $T_\mathrm{i}$, well below the stability limit of silicates as found in most of our models results (i) from slow particle growth in the continuously diluting and cooling gas or (ii) from an onset of silicate dust condensation at those low temperatures. In the first case silicate dust formation commences at higher temperature, but dust grows only slowly in the outflowing gas such that the dust density peaks at large distances, i.e. low temperature, once sufficient dust has formed for rapid acceleration of the dusty gas which then results in a rapid density decrease with increasing distance from the star. In the second case one possibly has an onset of dust formation at rather low temperatures \citep[e.g.][]{Gai13}.  More detailed models considering dust particle growth are required for deciding between these possibilities, which, however, are out of the scope of this paper. 

For all our sample stars a considerable iron content of the silicate material is required to obtain a satisfactory spectrum fit. The iron content is determined by interpolation between the Fe/(Fe+Mg) ratio of our four amorphous silicate samples. The fit quality of the peak position of the two silicate bands and of the short-wavelengths flank of the 10 $\mu$m feature allows to fix the relative contribution of the two components between which one has to interpolate sufficiently well. It appears desirable however to determine optical data also for iron contents intermediate between that of our four samples. This would enable a more accurate determination of the iron content than from our admittedly rather coarse grid.     

A fit of the spectra with the iron-free silicate Si-film1 and also with the iron-free silicate of \citet{Jae03} was not possible. It remains to check whether this is a peculiarity of our selected sample or if this looks different if larger numbers of objects are studied. 

Table \ref{TabModFit} shows the dust masses required for an optimum fit in the range from the inner edge $R_\mathrm{i}$ where a dust species condenses to the outer radius $R_\mathrm{a}$ at $10^4R_*$ assumed in the model calculation. More instructive would be to convert this into a dust mass-loss rate or to the fraction of the dust forming elements condensed into dust. Unfortunately this requires knowledge of the velocity structure of the stellar wind. The assumed variation $\propto r^{-2}$  of the density for the model calculation requires a constant velocity, which is not realistic since the radiation pressure on dust accelerates the outflow increasingly with progress of dust condensation. The simplification of a constant velocity is acceptable for calculating the dust mass in optically thin shells, but may be a problem for determining dust mass-loss rates. This is particularly a problem for the radius range where corundum but no other dust species exists, because in that range the velocity may be significantly lower than in the range where the dominant dust species, silicates and possibly iron, exist and accelerate the dusty gas. We also cannot simply use the observed outflow velocity as listed in Table \ref{TabStarParm} because this refers to the asymptotic value of the velocity at large distances. In order to give a crude estimate despite the uncertainty of such an approach we assume an average velocity in the dust condensation zone of $v=10$ km\,s$^{-1}$ for all stars and species (typically one half of the final velocity) and calculate a dust mass-loss rate from $\dot M_{\mathrm{du}}=M_{\mathrm{du}}\,v/(R_\mathrm{a}-R_\mathrm{i})$. The result is shown in Table \ref{TabMLR}. This can be compared to the gas mass-loss rate which shows that, as to be expected, the dust mass-loss rate is a fraction of about $10^{-2}$ to $<10^{-3}$ of the gas mass-loss rate. The accuracy of the estimated dust mass-loss rate is, however, low because the value of $\dot M_\mathrm{gas}$ is only known for $\mu$ Cep with some accuracy; in the other cases the values are only rough estimates.

\begin{table}
\caption{Estimated dust mass-loss rates.}
\begin{tabular}{l@{\hspace{.5cm}}rrrrrrr}
\hline
\hline
\noalign{\smallskip}
 & \multicolumn{2}{c}{Silicate} & \multicolumn{2}{c}{Corundum} & \multicolumn{2}{c}{Iron} 
\\[.2cm]
Object & \multicolumn{1}{c}{$\dot M_\mathrm{du}$} & \multicolumn{1}{c}{$f$} & \multicolumn{1}{c}{$\dot M_\mathrm{du}$} & \multicolumn{1}{c}{$f$} & \multicolumn{1}{c}{$\dot M_\mathrm{du}$} & \multicolumn{1}{c}{$f$} & \multicolumn{1}{c}{$\dot M_\mathrm{gas}$} \\
\noalign{\smallskip}
\hline
\noalign{\smallskip}
W Per     & 18.3 & 1.6   & 9.0 & 14.3 & 40.3 & 10.9 & 2100 \\
RW Cyg    & 10.0 & 0.57  & 7.4 & 7.7  & 10.2 &  1.8 & 3200 \\
RS Per    &  1.4 & 0.12  & 1.3 & 2.2  & 4.0  &  1.1 & 2000 \\
$\mu$ Cep &  2.3 & 0.083 & 1.3 & 0.84 & 9.1  & 1.03 & 5000 \\
\noalign{\smallskip}
\hline
\end{tabular}
\\
\tablecomments{$\dot M_\mathrm{du}$ is mass-loss rate of the dust species in units of $10^{-9}\,\rm M_{\sun}\,a^{-1}$, $f$ is estimated fraction of the key element for dust formation (Si for silicates, Al for corundum, Fe for iron) condensed into dust.}
\label{TabMLR}
\end{table}

We can even go a step further and calculate from $f=\dot M_\mathrm{du}/(A\,\epsilon \dot M_\mathrm{gas})$ a condensation fraction $f$ of the key elements (Si for silicates, Al for corundum, Fe for iron) for condensation into dust, where $A$ is the atomic weight of the chemical formula unit of the solid and $\epsilon$ the element abundance of the key element. The results are also shown in Table \ref{TabMLR}. In principle the fraction $f$ should satisfy $f<1$. This is satisfied for $\mu$ Cep, where the value of $\dot M_\mathrm{gas}$ is accurately known. In the other cases one obtains in part distinctively higher values. For corundum dust this may result from a lower than assumed value of $v$ in the inner range of the dust shell where only corundum dust exists, as mentioned above. For iron it may result from the fact that we used optical data for pure iron, while iron in space always is a nickel-iron alloy which may have different extinction properties. And also the estimated dust mass-loss rates as derived by \citet{Mau11} from the formula of \citet{Jur90} may be systematically too low. A more accurate modeling of the dust shell considering the growth of grains and the outflow dynamics is required to determine the condensation degrees for the dust species.

\section{Concluding remarks}

This paper studies the optical properties of iron-bearing amorphous silicates with oxygen to silicon ratios deviating from the stoichiometry of the usually considered silicates with olivine- and pyroxene-like compositions. We show that for four selected supergiants used as test-objects their observed mid-infrared emission as determined by ISO may be well explained with the presence of such kind of material.

We produce by a method of vapor-deposition on a cold substrate thin films of  truly amorphous silicates with different iron and magnesium contents and non-standard oxygen to silicon ratios that are within the range of values observed for presolar grains. The high degree of homogeneity and uniform thickness of the produced films achieved by our experimental method allows an accurate determination of the optical constants of the material. We report here on our results for four amorphous magne\-sium-iron silicates with non-standard stoichiometric composition: for the two end-member compositions of pure magnesium silicate and pure iron silicate, and for two intermediate values of the iron contents with magnesium-rich and iron-rich composition

We determine new optical constants for the four materials by IR spectroscopic transmittance measurements and by ellipsometry. From the combined data a model dielectric function is constructed that can be used to analyze circumstellar emission spectra. 

The new data are applied to a sample of four selected supergiants which are optically thin at $\lambda=10\ \mu$m but show substantial infrared excess emission. We aim to check how well we can fit observed ISO-spectra for these stars with synthetic spectra calculated for emission from a circumstellar dust shell enshrouding these stars. Since real stars generally seem to form also aluminium oxide dust if they form silicate dust, we include such dust in our model calculation. We also included a separate iron dust component, because such dust also seems to be present. The dust mass of each species and the temperature at the inner edge of the respective dust shell are determined from an automated fit between the synthetic spectrum from the radiative transfer model and the observations. To obtain a good fit we have to interpolate with respect to the Fe/(Mg+Fe) ratio between the four data sets. We found:

\begin{enumerate}

\item It is possible to obtain a good fit between model and observation with the kind of amorphous silicate with non-standard stoichiometry studied in this paper.

\item A good fit requires a significant iron content of the silicate material. 

\item The 10 $\mu$m and 18 $\mu$m features can be fitted simultaneously with our non-stoichiometric, amorphous, iron-bearing silicates.

\end{enumerate}
Though the obtained fits are not perfect, they are so close that the remaining deviations appear to be more likely due to shortcomings in the modeling (e.g. the assumption of CDE) than to not completely realistic optical constants. In particular it is not necessary to invoke additional dust species to reproduce the peak wavelengths and strengths of the two broad 10 $\mu$m and 18 $\mu$m amorphous silicate features.  

Also metallic iron dust particles seem to be formed in the outflow, which form separate particles not in physical contact with the silicate dust grains. They are required to explain the observed mid infrared excess between $\lambda=5.5$ $\mu$m and $\lambda=8$ $\mu$m, but a reasonable fit for the spectral region  $\lambda>9\ \mu$m dominated by the silicate emission can be obtained without such a component, if one leaves open the question on the origin of the flux excess at $\lambda<8\ \mu$m. 

For future analysis of much broader samples of stars  it would be desirable to have at our hands the optical data for a much finer grid of Fe/(Mg+Fe) ratios. The corresponding thin films are already produced and the raw data have already been acquired for this.  


\begin{acknowledgements}
We acknowledge the very constructive referee report which helped to improve the presentation of the results. We greatly acknowledge S. Hony for making available to us a set of reduced ISO spectra for our sample stars. We express our gratitude to Dr.~Theiss for making available for us the SCOUT software. We are thankful to Dr. S. Wetzel for his technical and scientific supports. We thank Hans-Werner Becker and Detlef Rogalla for Rutherford Backscattering analysis. Our project has been supported by the `Deutsche Forschungs\-gemeinschaft (DFG)' under the Priority Program SPP1385. This work is (partly) performed at and supported by RUBION, central unit of the Ruhr-Universit\"at Bochum. This research has made use of NASA's Astrophysics Data System.
\end{acknowledgements}

\begin{appendix}

\section{Fitting of synthetic spectra}

\label{AppRad}

\subsection{Model Assumptions}

The synthetic spectrum of a dust shell is calculated by taking advantage of the assumption of an optically thin dust shell. This allows a very rapid calculation of a model spectrum and this in turn enables computing large sets of such models for obtaining optimized fits between model spectra and observations. The model is calculated as follows:

1. For the emission from the central star we assume a black body spectral energy distribution with effective temperature $T_\mathrm{eff}$. Alternatively a read-in spectral energy distribution can be used. 

2. The dust temperature $T_\mathrm{d}$ is determined from radiative equilibrium between absorption and emission of a dust grain
\begin{equation}
\int_0^\infty \kappa^{\rm(abs)}_\nu B_\nu(T_\mathrm{eff})W(r)\,{\rm e}^{-\tau_\nu}\,{\rm d}\nu=\int_0^\infty 
\kappa^{\rm(abs)}_\nu B_\nu(T_\mathrm{d})\,{\rm d}\nu\,,
\label{TempDu}
\end{equation}
where $B_\nu$ is the Kirchhoff-Planck function, $\kappa^{\rm(abs)}_\nu$ the mass absorption coefficient of the dust material, and $W(r)$ is the geometric dilution factor of the stellar radiation field
\begin{equation}
W={1\over2}\left(1-\sqrt{1-{R_*^2\over r^2}}\right)\,
\end{equation}
where $R_*$ is the stellar radius. Each dust species has its own dust temperature $T_\mathrm{d}$. The quantity 
\begin{equation}
\tau_\nu=\sum_j\int_{R_+}^r\varrho_j\left(\kappa^{\rm(abs)}_{j,\nu}+\kappa^{\rm(sca)}_{j,\nu}\right)\,{\rm d}r'
\end{equation}
is the optical depth between the stellar surface and the radial distance $r$ of the dust grain from the stellar centre. The density $\varrho_j$ is the mass density of the dust species $j$ and the summation is over all dust species present. Our assumption of an optical thin dust shell mainly refers to the mid to far infrared wavelength for which we compare the emission from the dust shell with observations. The heating of the dust grains is due to absorption of stellar radiation which peaks around $1\dots1.5\ \mu$m. Since the extinction by dust grains strongly raises from the near infrared to the ultraviolet, the dust shell may be optically thin in the mid to far infrared wavelength even if its optically thickness is not small in the optical spectral region. For this reason we retain the factor $\exp(-\tau_\nu)$ despite our assumption of an optically thin dust shell. 

3. For each dust species it is assumed that the dust is spherically symmetric distributed around a central star and extends from some inner radius $R_i$ to an outer radius $R_a$. The radial distribution of the dust density $\varrho$ is assumed to follow a power law $r^{-p}$ where $p$ is generally assumed to equal $p=2$ which corresponds to a stellar wind with constant outflow velocity. The inner radius, $R_i$, is determined such that the dust temperature of the species takes a prescribed value $T_\mathrm{c}$ at this radius.  The outer radius, $R_\mathrm{a}$, is prescribed.

4. The observed spectral energy flux from the object observed at Earth is
\begin{align}
F_\nu=&{\rm e}^{-\tau^{\rm ISM}_\nu}{4\pi\over D^2}\ 
\left\{\frac14B_\nu\left(T_{\rm eff}\right)\,{\rm e}^{-\tau_\nu} R_*^2
+\sum_j\int\limits_{R_\mathrm{i}}^{R_{\rm a}}B_\nu\left(T_{{\rm d},j}\right)\,\varrho_j^{\phantom{j}}\kappa_{j,\nu}^{(\rm abs)}\,r^2\,{\rm d}r\right\} \,,
\label{SpectThin}
\end{align}
where $D$ is the distance of the object from Earth and $\tau^{\rm ISM}_\nu$ the optical depth of the ISM along the sightline. The first term in curly brackets is the contribution from the star and the second that of the dust shell. 

For optically thin dust shells the dust temperature distribution calculated from equation~(\ref{TempDu}) is spherically symmetric, even if the dust distribution is not. Therefore we introduced in the emission term spherical coordinates. More generally the corresponding integral is a volume integral
\begin{displaymath}
\int B_\nu\left(T_{{\rm d},j}\right)\,\varrho_j^{\phantom{j}}\kappa_{j,\nu}^{(\rm abs)}\,{\rm d}V\,,
\end{displaymath}
where one may decompose the integration domain into spherical shells and perform the angular integrations over each shell separately. Then one can write the volume integral equally well in terms of the average mass densities of the shells. Hence, we have to interpret $\varrho$ in equation~(\ref{SpectThin}) as the mass density averaged over concentric shells around the star. In this sense equation~(\ref{SpectThin}) also holds if the real dust distribution is not strictly spherically symmetric.  

5. The model allows to consider an arbitrary number of dust species. The mass absorption and scattering coefficients are calculated from the complex index of refraction. For inhomogeneously composed particles an effective complex index of refraction according to the Bruggeman mixing rule \citep[cf.][]{Ber95} can be calculated. The calculation can be done for either spherical particles (Mie theory) or for small ellipsoidal particles \citep{Boh83} with a numerically specified distribution of axis ratios and radii, for a continuous distribution of ellipsoids \citep[CDE, see][]{Boh83}, or for small cubes \citep{Fuc75}.

\subsection{Model Parameter}

In order to calculate a model for the spectral energy distribution of a given object with observed infrared spectrum one has to specify the luminosity $L$ and effective temperature $T_\mathrm{eff}$ of the star. The stellar radius follows from the standard relation $4\pi R_*^2\sigma T_\mathrm{eff}^4=L$. Further one has to specify the dust species considered in the model, the corresponding optical properties, and the particle shape and size distribution. For all species $\kappa_{j,\nu}^{(\rm abs)}$ is calculated at the specified wavelength grid. For the dust distributions of all species $j$  one fixes the power $p$, the outer radius of the dust shell, $R_\mathrm{a}$, and the total dust masses, $M_j$, and condensation temperatures, $T_{j,\mathrm{c}}$, of the dust species.

\subsection{Model Calculation}

For a given set of parameters the radial variation of dust temperature for each species is calculated according to equation~(\ref{TempDu}) for a small set of assumed values for $R_i$ and the value of $R_i$ corresponding to the specified value of $T_{j,\mathrm{c}}$ is determined by backward interpolation. Then the dust temperatures for $r>R_i$ are calculated and the spectral energy distribution is calculated according to equation~(\ref{SpectThin}). 
 
\subsection{Optimization}

For given dust properties a model depends on the two parameters $L$, $T_\mathrm{eff}$ and the $2J$ parameters $M_j$, $T_{j,\mathrm{c}}$ where $J$ is the number of dust species. The parameters $L$, $T_\mathrm{eff}$ are derived from astronomical observations and are given quantities for an object.\footnote{%
If spectral data in the optical and near IR are given, $T_\mathrm{eff}$ may also be determined by the optimization procedure. 
}
The parameters $M_j$, $T_{j,\mathrm{c}}$ of the dust shell are unknown and have to be determined by fitting the observed spectrum with the model spectrum. 

The goodness of fit between the synthetic and the observed spectrum is checked by calculating the quality function
\begin{equation}
\chi^2=\sum_l\left(\cal{N}F_\nu^{\rm obs}-F_\nu^{\rm calc}\over F_\nu^{\rm calc}\right)^2\,.
\end{equation}
For calculating $\chi^2$ not all frequency points used for calculating the spectrum are taken into account, but only those where no strong molecular features from the underlying star are seen in the observed spectrum. The quantity $\cal{N}$ is a normalization factor which accounts for the circumstance that usually the distance $D$ is not accurately known such that to enable a comparison of the two spectra one has to apply a shift to one of them to put them one upon another. This factor is determined such as to minimize $\chi^2$ as the solution of 
\begin{equation}
{\partial\,\chi^2\over\,\partial{\cal{N}}}=0
\end{equation}
for $\cal{N}$. The value of $\chi^2$ calculated with this normalization is taken as the quality function. Then we vary the $2J$ parameters dust masses $M_j$ and condensation temperatures $T_{j,\mathrm{c}}$ and search for the parameter combination for which $\chi^2$ takes the lowest value. This parameter set is then considered as the best-fit model for the dust shell.

Such nonlinear minimization problems are notoriously difficult to solve. Here the minimization of $\chi^2$ is done by applying a genetic algorithm that usually allows an efficient minimum search even if many parameters are involved. We use the special variant of such an algorithm described in \citet{Cha95} because this worked quite well in other projects where we used this method. 

The genetic algorithm operates with discrete values of the model parameters. The optimized parameter set resulting from this method therefore is slightly off from the true minimum. Instead of improving the accuracy of the genetic algorithm by using a finer resolution of the variable space we perform a final refinement step by minimization with a gradient method. This converges rapidly because one starts already from a position in variable space very close to the searched-for minimum.

\end{appendix}
\software{SCOUT{\citep{Scout2011}}},\software{PIKAIA{\citep{Cha95}}}


\end{document}